\documentclass{article}
\usepackage{amsmath}
\usepackage{amssymb}
\usepackage{array}
\usepackage{authblk}
\usepackage{booktabs}
\usepackage{caption}
\usepackage{comment}
\usepackage{csquotes}
\usepackage{dsfont}
\usepackage[export]{adjustbox}
\usepackage{geometry}
\usepackage{graphicx}
\usepackage{hhline}
\usepackage{lineno}
\usepackage{lscape}
\usepackage{mathtools}
\usepackage{multirow}
\usepackage[super]{natbib}
\usepackage{soul}
\usepackage{setspace}
\usepackage{subcaption}
\usepackage{textcomp}
\usepackage{xcolor}
\newcolumntype{M}[1]{>{\centering\arraybackslash}m{#1}}
\numberwithin{equation}{section}

	\title{A first passage model of intravitreal drug delivery and residence time, in relation to ocular geometry, individual variability, and injection location}

	\author[1]{Patricia Lamirande}
 	\author[1]{Eamonn A Gaffney}
 	\author[2]{Michael Gertz}
 	\author[1]{Philip K Maini}
        \author[1]{Jessica~R~Crawshaw}
        \author[2]{Antonello Caruso}

	\affil[1]{Wolfson Centre for Mathematical Biology, Mathematical Institute, Andrew Wiles Building, University of Oxford, Oxford, United Kingdom}
 	\affil[2]{Pharmaceutical Sciences, Roche Innovation Center Basel, Roche Pharma Research and Early Development, Basel, Switzerland}

\begin{document}
\maketitle
\footnotetext[1]{Corresponding author: lamirande@maths.ox.ac.uk}

\vspace{1cm}
\section*{Structured abstract}
\textbf{Purpose:} Standard of care for various retinal diseases involves recurrent intravitreal injections. This motivates mathematical modelling efforts to identify influential factors for drug residence time, aiming to minimise administration frequency. We sought to describe the vitreal diffusion of therapeutics in nonclinical species used during drug development assessments. In human eyes, we investigated the impact of variability in vitreous cavity size and eccentricity, and in injection location, on drug elimination.\\
\\
\textbf{Methods:} Using a first passage time approach, we modelled the transport-controlled distribution of two standard therapeutic protein formats (Fab and IgG) and elimination through anterior and posterior pathways. Detailed anatomical 3D geometries of mouse, rat, rabbit, cynomolgus monkey, and human eyes were constructed using ocular images and biometry datasets. A scaling relationship was derived for comparison with experimental ocular half-lives. \\
\\
\textbf{Results:} Model simulations revealed a dependence of residence time on ocular size and injection location. Delivery to the posterior vitreous resulted in increased vitreal half-life and retinal permeation. Inter-individual variability in human eyes had a significant influence on residence time (half-life range of 5-7 days), showing a strong correlation to axial length and vitreal volume. Anterior exit was the predominant route of drug elimination. Contribution of the posterior pathway displayed a small ($\sim$ 3\%) difference between protein formats, but varied between species (10-30\%).\\ 
\\
\textbf{Conclusions:} The modelling results suggest that experimental variability in ocular half-life is partially attributed to anatomical differences and injection site location. Simulations further suggest a potential role of the posterior pathway permeability in determining species differences in ocular pharmacokinetics.

\newpage
\doublespacing

\section*{Introduction}
The eye is a complex organ that varies significantly in size and shape between different species. In the human eye, individual variations of size and shape are common and can cause various vision conditions. With emmetropia describing the absence of refractive error, myopia is generally characterised by an elongated eye~\cite{Matsumura2019Update}, with a larger axial length (AL) compared to an emmetropic eye~\cite{Atchison2004Eye}, while hypermetropia is often associated with a shorter AL~\cite{Strang1998Hyperopia}.
\\
\\
Approximately one in three people have some form of disease-induced vision impairment by the age of~65~\cite{Quillen1999Common}. A common cause of vision loss among the elderly is age-related macular degeneration (AMD), a progressive disease characterised by damage to the macula~\cite{Quillen1999Common}. The wet form of AMD is characterised by upregulation of the vascular endothelial growth factor (VEGF), an angiogenic protein~\cite{Mitchell2018Age} that induces pathological neovascular growth leading to retinal damage~\cite{Chappelow2008Neovascular, Shweiki1992Vascular}.
\\
\\
Among treatment options for wet AMD are intravitreal (IVT) injections of protein therapeutics that bind to VEGF to inhibit its function~\cite{Mitchell2018Age}. Two standard-of-care therapeutics are ranibizumab, a monoclonal antibody fragment (Fab), and aflibercept, an Fc-fusion protein, with reported hydrodynamic radii ($R_h$) of 3.0 and 5.2 nm, respectively~\cite{Caruso2020Ocular, Shatz2016Contribution}. The latter is comparable to the macromolecular size of bevacizumab (5.0 nm~\cite{Caruso2020Ocular}), a monoclonal full-length IgG1 antibody~\cite{Meyer2011Preclinical} that is used off-label for the treatment of choroidal neovascularization. These antibodies are administered through IVT injections, and frequent administrations were shown to improve visual acuity outcome in the majority of patients~\cite{Rosenfeld2006Ranibizumab, Brown2006Ranibizumab, Bernoff2018Boundary}. However, IVT drugs exhibit suboptimal drug retention, with clinical studies reporting the ocular half-lives of less than 10 days~\cite{Krohne2008Intraocular, Krohne2012Intraocular}.
\\
\\
IVT injections in the human eye are not targeting a specific injection site within the vitreous chamber~\cite{Xing2014Survey}. Broad guidelines have been specified, for example the needle tip should be inserted more than 6 mm aiming at the centre of the eye~\cite{Aiello2004Evolving} with the bevel facing upward~\cite{Peyman2009Intravitreal}, which provides less than a precise target for the injection site. It is also advised to deliver the dosing formulation gently into the vitreous cavity with a slow injection, in order to avoid jet formation and excessive cavitary flow~\cite{Aiello2004Evolving, Peyman2009Intravitreal}. In general, procedures are specified with focus on preventing mechanical damage and infections, with seemingly less attention paid to the potential impact on drug absorption or ocular residence time~\cite{Aiello2004Evolving, Fagan2013Intravitreal}. However, chronic treatment places a burden on patients and healthcare systems in terms of resources and procedural risks~\cite{Chong2016Ranibizumab, Doshi2011Intravitreal}. Moreover, it is estimated that the majority of the injected drug is eliminated through the anterior pathway via aqueous humour turnover, with a small proportion permeating the retina (posterior elimination pathway), despite being the target site of action~\cite{delAmo2017Pharmacokinetic,Hutton2017Ocular}. This motivates investigation of drug residence time in the eye due to differences in eye shape and size, drug hydrodynamic radii, and injection locations.
\\
\\
The translation of results across nonclinical species and patients is crucial for the effective design and characterization of drug candidates. Species commonly used in ocular pharmacokinetic~(PK) or pharmacodynamic~(PD) studies are the rabbit~\cite{Ameri2007Effects, Ahn2014Intraocular, Gaudreault2007Pharmacokinetics, Shatz2016Contribution, Bakri2007Pharmacokinetics, Christoforidis2011Pet, Kim2021Permeability}, cynomolgus monkey~\cite{Niwa2015Ranibizumab, Gaudreault2005Preclinical, Miyake2010Pharmacokinetics}, and rat~\cite{Lu2009Intravitreal, Gal2016Bevacizumab}, with fewer studies reported in pig~\cite{Stricker2016Miniature, Shrader2018Gottingen, Kelley2022Generation} and mice~\cite{Bussing2023Pharmacokinetics, Schlichtenbrede2009Toxicity}. Previous studies have demonstrated the role of diffusion in determining the vitreal elimination rate of IVT macromolecules, with PK experiments showing a dependence of ocular half-life on both the molecular size and eye size~\cite{Caruso2020Ocular, Shatz2016Contribution, Crowell2019Influence}. Nevertheless, this relationship appears to be species-specific and influenced by other factors besides diffusion distance. In fact, the experimental half-life obtained for a given molecule or molecular size is larger in pigs than in humans, in rabbits than in monkeys, although the respective vitreous volumes are smaller. Caruso et al.~\cite{Caruso2020Ocular} and Crowell et al.~\cite{Crowell2019Influence} postulated that the observed species-specific PK could result from differences in ocular shape and eccentricity or in the contribution of the posterior pathway to drug elimination, amongst other factors. This motivates modelling efforts in investigating these aspects to support the translational characterisation of novel drug candidates.
\\
\\
Scaling relationships between species have been previously established~\cite{Hutton2016Mechanistic} under the assumption of spherical vitreous chambers, implying simplified ocular anatomies. More anatomically faithful models are also available, such as the computational fluid dynamics works of Missel~\cite{Missel2012Simulating} and Lamminsalo et al.~\cite{Lamminsalo2018Extended, Lamminsalo2020Extended} that describe both the posterior and anterior segments in great detail. Besides the complexity and computational cost inherent to such models, the rate of egress of material from the vitreous remains the rate-limiting factor and main determinant of ocular PK, so studying the distribution within the vitreous cavity and at its interfaces in more detail is key.
\\
\\
As defined in the random walk field, the first passage time is a random variable describing how long it takes for a random walker to reach a given target site~\cite{Berg1977Physics, Benichou2014First}. Its expected value is called the mean first passage time (MFPT). The MFPT has been described as an effective measure of diffusive transport~\cite{Redner2001Guide, Gardiner2009Stochastic, Bressloff2013Stochastic}. Recent applications in mathematical biology have been diverse, including applications to animal movement in heterogeneous landscapes~\cite{Mckenzie2009First}, receptors in the synaptic membrane~\cite{Holcman2004Escape}, and drug molecules crossing the mucus-epithelium interface~\cite{Newby2018Technological}. In ocular modelling, an approximation of the MFPT was previously used to define the vitreal diffusion time, the average time for a particle to diffuse from the centre of a sphere to its surface, with the vitreous chamber modelled as a sphere in Hutton-Smith et al.~\cite{Hutton2016Mechanistic}. However, the first-passage problem was not explicitly solved.  In the present MFPT modelling framework, we considered isotropic diffusion and the absence of convection in the distribution of the injected ocular drug, justified by the slow injection of the drug and by previous experiments supporting the absence of flow in the vitreous chamber~\cite{Maurice1957Exchange, Moseley1984Routes, Gaul1986Measurement, Maurice1987Flow, Araie1991Loss}. The MFPT is a measure of the residence time (which does not depend on initial drug concentration), and its framework can be extended to quantify the drug elimination using the distribution of exits, the multidimensional analogue of the splitting probabilities~\cite{Karlin1981Second, Redner2001Guide, Gardiner2009Stochastic}. This can be used to quantify the amount of drug leaving through each elimination pathway and their relative importance.  \\
\\
This research aims to investigate the influence of ocular size and shape, inter-individual variability, drug molecular features, and injection location on the vitreal kinetics and residence time of IVT macromolecules. We further aim to study whether inter-species differences in vitreous chamber geometry may explain the different pharmacokinetics observed experimentally. To this end, we develop a mathematical model of vitreal drug diffusion based on the first passage time methodology, deriving equations for the MFPT, the conditional MFPT and the drug elimination distribution for two standard molecular formats of IVT therapeutics, namely Fab and IgG. Using literature datasets to construct realistic 3D geometries, we model human emmetropic, myopic, and hypermetropic eyes for studying the influence of ocular shape on drug retention. We also model the anatomy of the vitreous chamber in the mouse, rat, rabbit, and cynomolgus monkey, aiming to improve the translational understanding of PK in support of drug discovery and development. In order to assess the influence of spatial parameters, we compare the residence time in the different vitreous chamber geometries and derive a scaling relationship between the MFPT, vitreal volume and axial length (AL). We also assess the dependence of retinal absorption on the spatial parameters describing the ocular geometries and on the site of injection within the vitreous cavity, and identify the dominant elimination pathway kinetics using the conditional MFPT.\\
\\

\section*{Methods}
\subsection*{Ocular geometry}
\begin{figure}
        \centering
        \begin{subfigure}{\linewidth}
            \includegraphics[width=\linewidth]{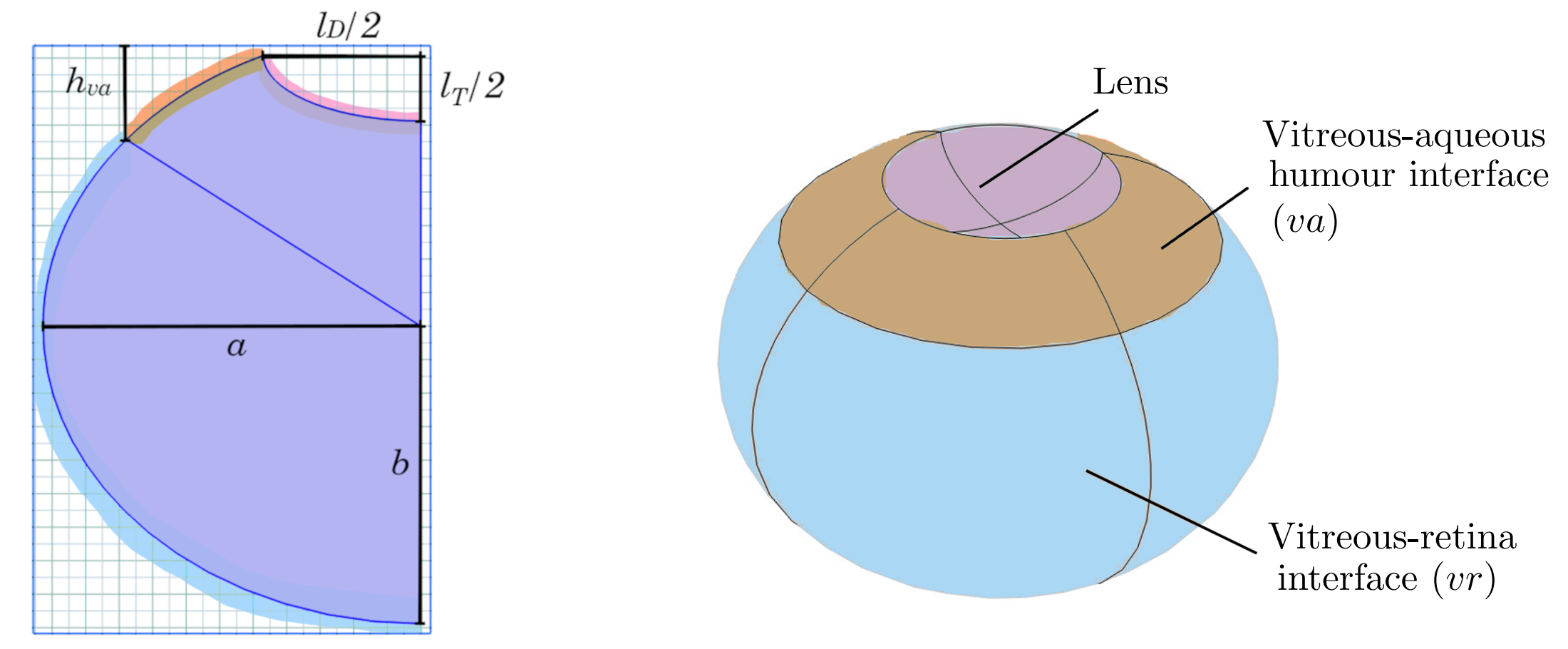}
        \end{subfigure}
        \caption{Plane geometry used in the axial rotation to define the 3D ocular model for the human eye, where $a$ and $b$ are the semi-major and semi-minor axis of the vitreous chamber ellipse, $l_D$ and $l_T$ are the lens diameter and thickness, and $h_{va}$ is the height of the vitreous-aqueous humour interface. The vitreous-lens interface is identified in pink, the vitreous-aqueous humour interface in orange, and the vitreous-retina interface in blue. Parameters not shown: $l_p$, the proportion of the lens thickness within the vitreous chamber ellipse, and $V_{vit}$ and $A_{ret}$: the volume of the vitreous humour and the area of the retinal surface.}
        \label{fig:plane_geom}
    \end{figure}

\begin{figure}
    \centering
    \includegraphics[width=\linewidth]{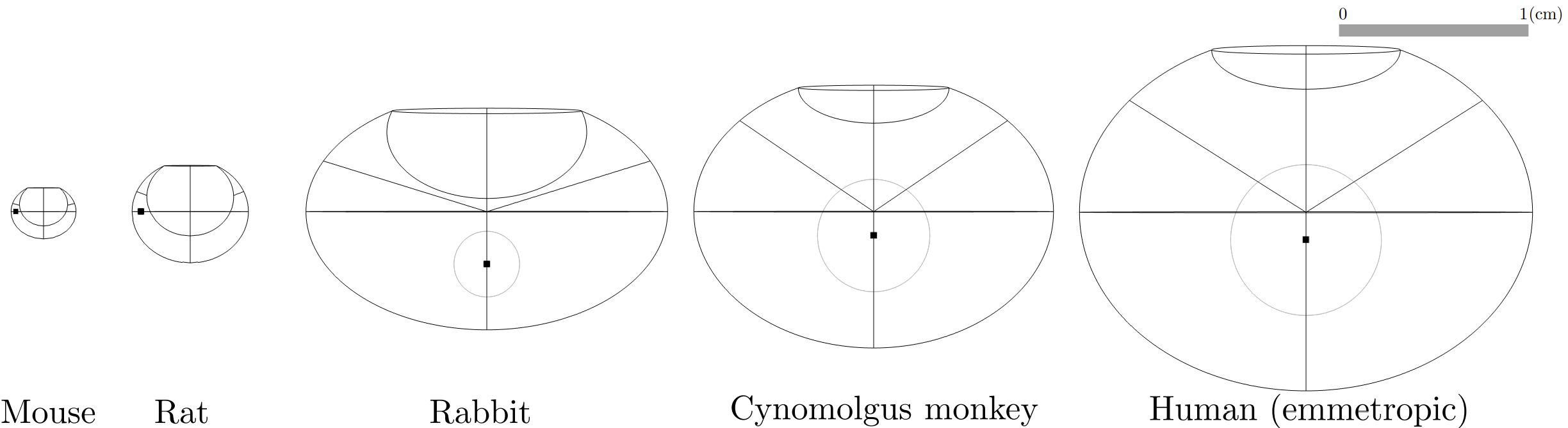}
    \caption{Cross-sections of the ocular geometries built using parameters in Table~\ref{tab:params_lit_review}, and based on Figure~\ref{fig:plane_geom}, in relative scale. The black bullet represents the injection point $P_m$, and the grey circle in the rabbit, cynomolgus monkey, and human eye models represents the injection location region considered in the ocular half-life ($t_{1/2}$) analysis.}
    \label{fig:geometries_uptoscale}
\end{figure}

\begin{figure}
    \centering
    \includegraphics[width=\linewidth]{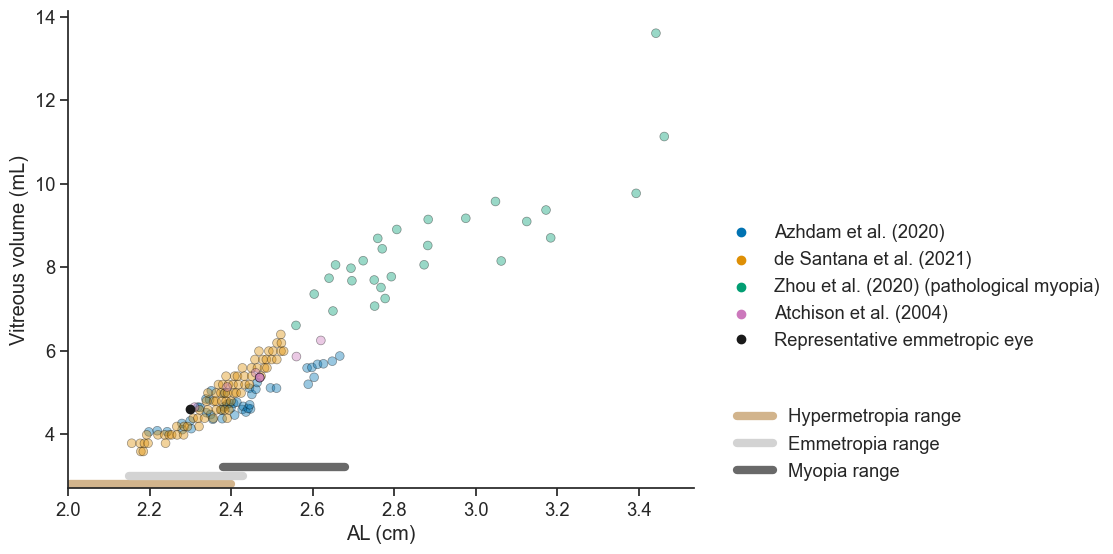}
    \caption{Literature measurements of vitreous volume and axial length (AL) in 155 human eyes, used to build the ensemble of human eye models. The hypermetropia, emmetropia, and myopia range are identified using definitions by Strang et al.~\cite{Strang1998Hyperopia} and Atchison et al.~\cite{Atchison2004Eye}. Black bullet: emmetropic human eye model corresponding to the geometry of Figure~\ref{fig:geometries_uptoscale}.}
    \label{fig:set_human_eyes_data}
\end{figure}

3D models of the vitreous chamber were built for human and relevant nonclinical species. The posterior cavity was assumed to be an oblate spheroid, obtained by the rotation of an ellipse around its minor axis, which was collinear with the optical axis. The lens protruding into the vitreous humour was similarly defined. The difference between the two determined the vitreous chamber (Figure~\ref{fig:plane_geom}). Three interfaces were defined, corresponding to the vitreous-lens, vitreous-aqueous humour, and vitreous-retina boundaries. Anatomically, the vitreous-retina interface corresponds to the surface covered by the inner limiting membrane (ILM) and delimited by the ora serrata, and the vitreous-aqueous humour interface corresponds to the zonular fibres and the space of Petit. The parameters used to build the 3D geometries are defined in Figure~\ref{fig:plane_geom}, and the details of the construction of the geometries are summarised in SI.B.
\\
\\
A literature search was conducted to collect the anatomical dimensions of human eyes as well as those of the cynomolgus monkey, rabbit, rat, and mouse. Insufficient anatomical information prevented inclusion of the pig or minipig among modelled species. A summary of the literature data is provided in Table~\ref{tab:params_lit_review}. The measurements were performed by various methods, including magnetic resonance imaging (MRI), optical coherence tomography, ultrasound biometry, Scheimpflug photography, and direct measurements of postmortem fixed eyes. The parameter values for the model geometries are also summarised in Table~\ref{tab:params_lit_review} and detailed in SI.B. A cross-section of the model for each species is displayed in relative scale in Figure~\ref{fig:geometries_uptoscale}. We verified the anatomical accuracy of the geometries by comparison with experimental measurements of vitreous volumes and retinal areas ($V_{vit}$ and $A_{ret}$ in Table~\ref{tab:params_lit_review}), and with in vivo MRI images obtained from the literature~\cite{Tkatchenko2010Analysis, Chui2012Refractive, Sawada2002Magnetic, Short2008Safety, Atchison2004Eye}.
\\
\\
To investigate their influence on ocular PK, different injection site locations within the vitreous chamber were considered. Under the assumption that IVT injections target the central vitreous, equidistantly from the retina and the posterior surface of the lens, a region of interest was defined around the midpoint ($P_m$) of the vitreous chamber depth. For simplicity, a sphere of diameter ($b-l_T/2$) was defined (Figure~\ref{fig:geometries_uptoscale}). The resulting volume encompasses the vitreous core, arguably representing a conservative estimate for the possible locations of IVT delivery. This injection region was employed to assess the impact on ocular half-life, $t_{1/2}$, in human, cynomolgus monkey, and rabbit eyes. In rodents, the lens occupies a significant portion of the posterior cavity~\cite{Tkatchenko2010Analysis, Chui2012Refractive}, giving the vitreous chamber a distinctive crescent shape, for which it is more difficult to define the centre and reach it with an injection needle. Therefore, in the rat and mouse model, $P_m$ was defined as the midpoint between the retina and the lens, along the vitreous diameter (Figure~\ref{fig:geometries_uptoscale}).\\
\\
Additionally, an ensemble of 155 human eye models was built based on axial length (AL) and vitreous volume measurements obtained from the literature (Figure~\ref{fig:set_human_eyes_data}). The measurements were collected by MRI~\cite{Atchison2004Eye, Zhou2020Quantitative}, optical biometry and vitrectomy~\cite{deSantana2021Use} and CT scan~\cite{Azhdam2020Vivo}. The AL data collectively covers the range associated with hypermetropic, emmetropic, and myopic eyes~\cite{Atchison2004Eye, Strang1998Hyperopia}, and includes data for pathological myopia, described as a refractive error of -8 diopter or lower~\cite{Zhou2020Quantitative, Morgan2012Myopia}. Using the vitreous volume and AL, the eye geometries were constructed assuming a constant lens thickness and anterior chamber depth. The reader is referred to SI.B for further details.\\
\\

\begin{landscape}
\begin{table}[]
\begin{spacing}{1.5}
\small
    \centering
    \begin{tabular}{|p{1.6cm} | p{4cm}  p{1.5cm} | p{3.5cm}  p{1.5cm} | p{3cm} p{1.5cm} |}
    \hline 
         Parameter & Mouse & &  Rat & & Rabbit  &  \\ 
         & Literature Value & Model Value & Literature Value & Model Value & Literature Value & Model Value \\ \hline
         $a$ (cm) & 0.161 - 0.163 \cite{Tkatchenko2010Analysis} & 0.1618 & 0.286 - 0.293 \cite{Hughes1979Schematic} & 0.2895 & 0.88 - 0.92 \cite{Sawada2002Magnetic} & 0.90\\
         $b$ (cm) & 0.127 - 0.139 \cite{Schmucker2004Vivo} & 0.1355 & 0.253 - 0.257 \cite{Hughes1979Schematic} & 0.255 & 0.566 - 0.611 \cite{Liu1998Twenty} & 0.588 \\
         $l_D$ (cm) & 0.223 - 0.245 \cite{Pan2023Age} & 0.240 & 0.423 - 0.51 \cite{Massof1972Revision, Hughes1979Schematic, Pe'er1996Epithelial} & 0.432 & 0.971 -  1.02 \cite{Werner2006Experimental} & 0.995\\
         $l_T$ (cm) & 0.197 - 0.241 \cite{Pan2023Age, Tkatchenko2010Analysis, Schmucker2004Vivo} & 0.216 & 0.371 - 0.457 \cite{Massof1972Revision, Hughes1979Schematic, Lozano2013Development} & 0.387 & 0.606 - 0.697 \cite{Atsumi2013Comparative, Liu1998Twenty} & 0.66\\
         $l_p$ (\%) & & 99 & & 99 &  & 66 \\ 
         $h_{va}$ (cm) & &  0.05 & & 0.07 & &  0.238 \\ 
         $V_\text{vit}$ (ml) & 0.0044 - 0.012 \cite{Kaplan2010Vitreous, Clough2003Anterior, Lin2022Protocol, Schlichtenbrede2009Toxicity} &  0.00842 & 0.0505 - 0.0543 \cite{Sha2006Postnatal} & 0.0518 & 1.15 - 1.80 \cite{delAmo2015Rabbit, Atsumi2013Comparative, Vezina2013Comparative} & 1.71\\
         $A_\text{ret}$ (cm$^2$)& 0.134 - 0.190 \cite{Jeon1998Major, Lyubarsky2004Candelas, Drager1981Ganglion} & 0.188 & 0.65 - 0.8 \cite{Mayhew1997Photoreceptor, Baden2020Understanding} & 0.667 & 4 - 6 \cite{Reichenbach1991Development} &  5.44 \\ \hline

         Parameter & Cynomolgus Monkey & &  Human & &   &  \\ 
         & Literature Value & Model Value & Literature Value & Model Value & & \\ \hline
         $a$ (cm) & 0.855 - 0.915* &  0.895 & 1.03 - 1.25 \cite{Atchison2004Eye, Azhdam2020Vivo, deSantana2021Use} &  1.1275 & & \\
         $b$ (cm) & 0.623 - 0.733 \cite{Choi2021NormativeCyno, Lapuerta1995Four} & 0.678 & 0.816 - 1.07 \cite{Atchison2004Eye, Koretz2004Scheimpflug, Azhdam2020Vivo, deSantana2021Use} &  0.889 & & \\
         $l_D$ (cm) & 0.73 - 0.79 \cite{Manns2007Optomechanical} & 0.75 & 0.88 - 0.985 \cite{Rosen2006Vitro, Werner2006Experimental, Manns2007Optomechanical} & 0.939 & & \\
         $l_T$ (cm) & 0.288 - 0.403 \cite{Lapuerta1995Four, Atsumi2013Comparative, Choi2021NormativeCyno} & 0.351 & 0.391 - 0.564 \cite{Koretz2004Scheimpflug, Rosen2006Vitro, Werner2006Experimental} &  0.3909 & & \\
         $l_p$ (\%) & & 50 & & 50 & & \\ 
         $h_{va}$ (cm) & &  0.163  & &  0.251 & & \\
         $V_\text{vit}$ & 2.0 - 2.3 \cite{Atsumi2013Comparative} & 2.20 & 3.58 - 6.38 \cite{Azhdam2020Vivo, deSantana2021Use} &  4.60 & & \\
         $A_\text{ret}$ (cm$^2$) & 5.8 - 9.2 \cite{Wikler1990Photoreceptor} & 6.91 & 10.12 - 13.63 \cite{Panda1994Retinal, Curcio1990Topography, Nagra2017Determination} & 11.0 & & \\ \hline
    \end{tabular}
    \caption{Literature values of ocular geometry measures, and ocular model dimensions. The derivation of the model parameters is detailed in SI.B. (*):~Estimated from $V_{vit}$.}
    \label{tab:params_lit_review}
    \end{spacing}
\end{table}
\end{landscape}

\subsection*{Equations}
The equations have been derived by expanding the first passage time approach~\cite{Berg1993Random, Redner2001Guide, Bressloff2013Stochastic, Gardiner2009Stochastic, Delgado2015conditional} and applying it to vitreal transport.
\\

\subsubsection*{Mean first passage time}
The MFPT, $\tau(\boldsymbol{x_0})$, for a particle starting at $\boldsymbol{x_0}$, satisfies the following partial differential equation (PDE) and boundary conditions:
\begin{equation}
\begin{aligned}
    -D \, \nabla^2 \tau(\boldsymbol{x_0}) &= 1 \hspace{2.25cm} \text{ for } \boldsymbol{x_0} \in  \Omega,\\
    \nabla \tau(\boldsymbol{x_0}) \cdot \boldsymbol{n} &= 0 \hspace{2.25cm} \text{ for } \boldsymbol{x_0} \in \partial \Omega_{vl}, \\
    -D \, \nabla \tau(\boldsymbol{x_0}) \cdot \boldsymbol{n} &= \kappa_{va} \, \tau(\boldsymbol{x_0}) \hspace{1cm} \text{ for } \boldsymbol{x_0} \in \partial \Omega_{va},\\
    -D \, \nabla \tau(\boldsymbol{x_0}) \cdot \boldsymbol{n} &= \kappa_{vr} \, \tau(\boldsymbol{x_0}) \hspace{1cm} \text{ for } \boldsymbol{x_0} \in \partial \Omega_{vr},
    \label{eq:mfpt_pde}
\end{aligned}
\end{equation}
where $\boldsymbol{n}$ is the outward normal and with parameters defined in Table~\ref{tab:parameters_diff_perm_fab_igg}, for the vitreal region $\Omega$ and associated lens, anterior and retinal boundaries $\partial \Omega_{vl}$, $\partial \Omega_{va}$ and $\partial \Omega_{vr}$ respectively, as illustrated in Figure~\ref{fig:plane_geom}.
\\
\begin{table}[b]
\small
\begin{spacing}{1.5}
    \centering
    \begin{tabular}{ m{8.2cm}  m{3.6cm}  m{3.4cm}}
    \hline
        Drug-dependent parameters & Fab & IgG  \\ \hline
        Diffusion coefficient ($D$)  & $1.07 \times 10^{-6}$ cm²/s \cite{Caruso2020Ocular} & $0.64 \times 10^{-6}$ m²/s \cite{Caruso2020Ocular} \\
        Permeability of vitreous-aqueous humour interface ($k_{va}$) & $1.91 \times 10^{-5}$ cm/s \cite{Hutton2018Modelling} & $0.874 \times 10^{-5}$ cm/s \cite{Hutton2018Modelling} \\
        Permeability of vitreous-retina interface ($k_{vr}$) & $1.81 \times 10^{-7}$ cm/s \cite{Hutton2018Modelling} & $1.19 \times 10^{-7}$ cm/s \cite{Hutton2018Modelling} \\ \hline
    \end{tabular}
    \caption{\raggedright Definition of the drug-dependent parameters for the Fab and IgG molecular formats.}
    \label{tab:parameters_diff_perm_fab_igg}
\end{spacing}
\end{table}
\\
The MFPT can be linked to the ocular half-life $t_{1/2}$, i.e. the time required for a quantity that is exponentially decaying to fall to one half of its initial value. In this context, it characterises the rate at which the drug is cleared from the vitreous chamber. Under the assumption that the drug concentration inside the vitreous is decreasing exponentially after an IVT injection (see SI.A), the MFPT corresponds to the inverse of the decay rate, and we obtain the relation:
\begin{equation}
    t_{1/2}(\boldsymbol{x_0}) = \tau(\boldsymbol{x_0}) \ln (2)
    \label{eq:link_mfpt_t1/2}
\end{equation}
for an injection at $\boldsymbol{x_0}$.\\

\subsubsection*{Drug elimination and conditional mean first passage time}
Let $\pi_{va}(\boldsymbol{x_0})$ be the proportion of drug leaving through the vitreous-aqueous humour interface ($\partial \Omega_{va}$), and $\pi_{vr}(\boldsymbol{x_0})$ the proportion leaving through the vitreous-retina interface ($\partial \Omega_{vr}$), where $\boldsymbol{x_0}$ is the initial position. Note that these are the only regions in the model where molecules can exit and, therefore, the proportions sum to 1. We define $T_{va}(\boldsymbol{x_0})$ as the MFPT conditional on drug molecules leaving through $\partial \Omega_{va}$, and $T_{vr}(\boldsymbol{x_0})$ the conditional MFPT for molecules leaving through $\partial \Omega_{vr}$, both functions of the molecules’ initial position~$\boldsymbol{x_0}$. The proportion of drug exiting through~$\partial \Omega_{vr}$, $\pi_{vr}(\boldsymbol{x_0})$, satisfies the PDE system~\cite{Bressloff2013Stochastic, Gardiner2009Stochastic, Delgado2015conditional}:
\begin{equation}
\begin{aligned}
    \nabla^2 \pi_{vr}(\boldsymbol{x_0}) &= 0 \hspace{3.75cm} \text{ for } \boldsymbol{x_0} \in  \Omega,\\
    \nabla \pi_{vr}(\boldsymbol{x_0}) \cdot \boldsymbol{n} &= 0 \hspace{3.75cm} \text{ for } \boldsymbol{x_0} \in \partial \Omega_{vl},\\
    -D \, \nabla \pi_{vr}(\boldsymbol{x_0}) \cdot \boldsymbol{n} &= \kappa_{va} \, \pi_{vr}(\boldsymbol{x_0}) \hspace{2.2cm} \text{ for } \boldsymbol{x_0} \in \partial \Omega_{va}, \\
    -D \, \nabla \pi_{vr}(\boldsymbol{x_0}) \cdot \boldsymbol{n} &= - \kappa_{vr} + \kappa_{vr} \,  \pi_{vr}(\boldsymbol{x_0}) \hspace{1cm} \text{ for } \boldsymbol{x_0} \in \partial \Omega_{vr},
    \label{eq:splitting_prob_vr}
\end{aligned}
\end{equation}
and $T_{vr}(\boldsymbol{x_0})$ satisfies \cite{Gardiner2009Stochastic, Delgado2015conditional}:
\begin{equation}
\begin{aligned}
    -D \, \nabla^2 [\pi_{vr}(\boldsymbol{x_0}) \, T_{vr}(\boldsymbol{x_0})] &= \pi_{vr}(\boldsymbol{x_0}) \hspace{3.05cm} \text{ for } \boldsymbol{x_0} \in  \Omega,\\
    \nabla [\pi_{vr}(\boldsymbol{x_0}) \, T_{vr}(\boldsymbol{x_0})] \cdot \boldsymbol{n} &= 0 \hspace{4.05cm} \text{ for } \boldsymbol{x_0} \in \partial \Omega_{vl},\\
    -D \, \nabla [\pi_{vr}(\boldsymbol{x_0}) \, T_{vr}(\boldsymbol{x_0})] \cdot \boldsymbol{n} &= \kappa_{va} \, [\pi_{vr}(\boldsymbol{x_0}) \, T_{vr}(\boldsymbol{x_0})] \hspace{1cm} \text{ for } \boldsymbol{x_0} \in \partial \Omega_{va}, \\
    -D \, \nabla [\pi_{vr}(\boldsymbol{x_0}) \, T_{vr}(\boldsymbol{x_0})] \cdot \boldsymbol{n} &= \kappa_{vr} \, [\pi_{vr}(\boldsymbol{x_0}) \, T_{vr}(\boldsymbol{x_0})] \hspace{1cm} \text{ for } \boldsymbol{x_0} \in \partial \Omega_{vr}.
\end{aligned}
\label{eq:pde_cond_mfpt_vr}
\end{equation}
The conditional MFPT for drug molecules leaving through the vitreous-aqueous humour, $T_{va}(\boldsymbol{x_0})$, was derived following the same method.
\\

\subsection*{Drug-dependent parameters}
The drug-dependent parameters present in the PDE systems~\eqref{eq:mfpt_pde} to \eqref{eq:pde_cond_mfpt_vr} were set using experimental measures and modelling results found in the literature. The value of the diffusion coefficient $D$ is well defined by experimental studies in the literature, and was set to $D = 1.07 \times 10^{-6}$~cm$^2$/s and $D = 0.64 \times 10^{-6}$~cm$^2$/s for a Fab and IgG molecular format, respectively~\cite{Caruso2020Ocular}.\\
\\
The permeability parameters $\kappa_{va}$ and $\kappa_{vr}$, for the vitreous-aqueous humour and vitreous-retina interface, respectively, were more difficult to determine. In addition to varying with the drug molecule size, the permeability parameters are reported to vary between different species~\cite{Loch2012Determination}. Previous estimates of these parameters for Fab and IgG molecules are summarised in Table~\ref{tab:perm_params_lit_review_fab_igg}. We begin our study by setting $\kappa_{va} = 1.91 \times 10^{-5}$~cm/s and $\kappa_{vr} = 1.81 \times 10^{-7}$~cm/s for a Fab molecule, identifying the vitreous-aqueous humour interface as equivalent to the hyaloid membrane in~\cite{Hutton2018Modelling}, and the permeability of the vitreous-retina interface to be the lowest permeability between the retinal pigment epithelium (RPE) and the inner limiting membrane (ILM). For the IgG molecule, we followed the same steps and set $\kappa_{va} = 0.874 \times 10^{-5}$ cm/s and $\kappa_{vr} = 1.19 \times 10^{-7}$ cm/s, using the estimations from~\cite{Hutton2018Modelling}.\\
\begin{table}[]
\begin{spacing}{1.5}
    \centering
    \begin{tabular}{|m{7.1cm} |m{3.3cm} |m{3.3cm}| m{1cm}|}
    \hline
         Permeability & Fab & IgG & Source \\
         \hline
         RPE permeability ($\times 10^{-7}$ cm/s) &  2.60 (1.36, 4.04) & 1.84 (1.08, 2.36) & \cite{Hutton2017Ocular}\\ 
         & 2.63 & & \cite{Hutton2018Theoretical}  \\
         & 2.48 (2.2, 5.35) & 2.31 (1.76, 2.98) & \cite{Hutton2018Modelling} \\ \hline
         
         ILM permeability ($\times 10^{-7}$ cm/s) & 1.88 (1.13, 2.81) & 1.7 (0.912, 2.32) & \cite{Hutton2017Ocular} \\
         & 1.89 & & \cite{Hutton2018Theoretical} \\
         & 1.81 (1.25, 2.44) & 1.19 (1.12, 1.55) & \cite{Hutton2018Modelling} \\ \hline
         
         Hyaloid membrane permeability ($\times 10^{-5}$ cm/s) & 1.91 (1.24, 3.92) & 0.874 (0.616, 1.42) & \cite{Hutton2018Modelling} \\ \hline
    \end{tabular}
    \caption{Estimated permeability parameters with 95\% confidence intervals (where provided) from different sources, all determined by fitting models to rabbit data.}
    \label{tab:perm_params_lit_review_fab_igg}
\end{spacing}
\end{table}

\subsection*{Numerical methods}
All geometries were built using COMSOL Multiphysics \textregistered~\cite{Comsol}. In constructing each geometry, a mesh for the finite element numerical method was also constructed within COMSOL, with sufficient grid resolution to ensure numerical convergence. This was tested by confirming that further mesh refinement had no impact on example results at the resolution of plotting presented.\\
\\
The equations were solved and the figures were generated using COMSOL Multiphysics \textregistered~software~\cite{Comsol}, using the implemented stationary solver. Regression lines were obtained with the Scikit-learn library~\cite{Scikit-learn} in Python.  \\

\subsection*{Global sensitivity analysis}
We performed a global sensitivity analysis to identify which geometrical parameters are most important to accurately determine when constructing computational ocular models. We varied the geometrical parameters $a$, $b$, $l_D$ and $l_T$ (see Figure~\ref{fig:plane_geom} for their definition) within the range of literature ocular values identified in Table~\ref{tab:params_lit_review} for each species, and $h_{va}$ within $\pm 10$ \% of its base value. We performed this sensitivity analysis for the human, cynomolgus monkey, and rabbit eye models, as the geometries for the rat and mouse models were only well defined for a subset of the parameter combinations. We analysed the effect of the geometrical parameters on the MFPT for an injection at $P_m$. This was implemented using the eFAST sensitivity method~\cite{Saltelli1999Quantitative, Marino2008Methodology}, with the python SAlib library~\cite{Iwanaga2022, Herman2017}, and with the choice of sampling parameters guided by the methodology proposed in the referenced sources. We set to 4 the number of harmonics to sum in the Fourier series decomposition and to 337 the number of samples to generate. The implementation of the sensitivity analysis sampling was validated by confirming that a dummy variable has sensitivity indices of around zero, demonstrating minimum sampling artefact~\cite{Marino2008Methodology}.
\\
\\

\section*{Results}

\subsection*{Mean first passage time}
\begin{figure}[h]
    \centering
    \begin{subfigure}{\linewidth}
        \includegraphics[width=\linewidth]{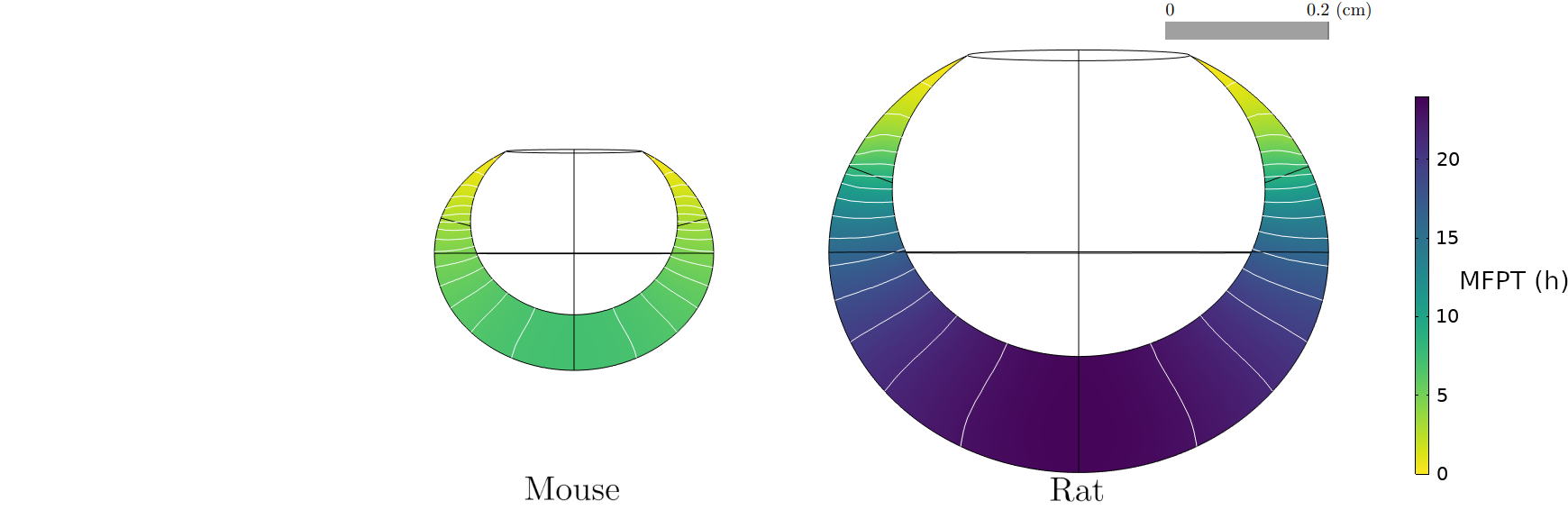}
    \end{subfigure}
    \\
    \vspace{1cm}
    \begin{subfigure}{\linewidth}
        \centering
        \includegraphics[width=\linewidth]{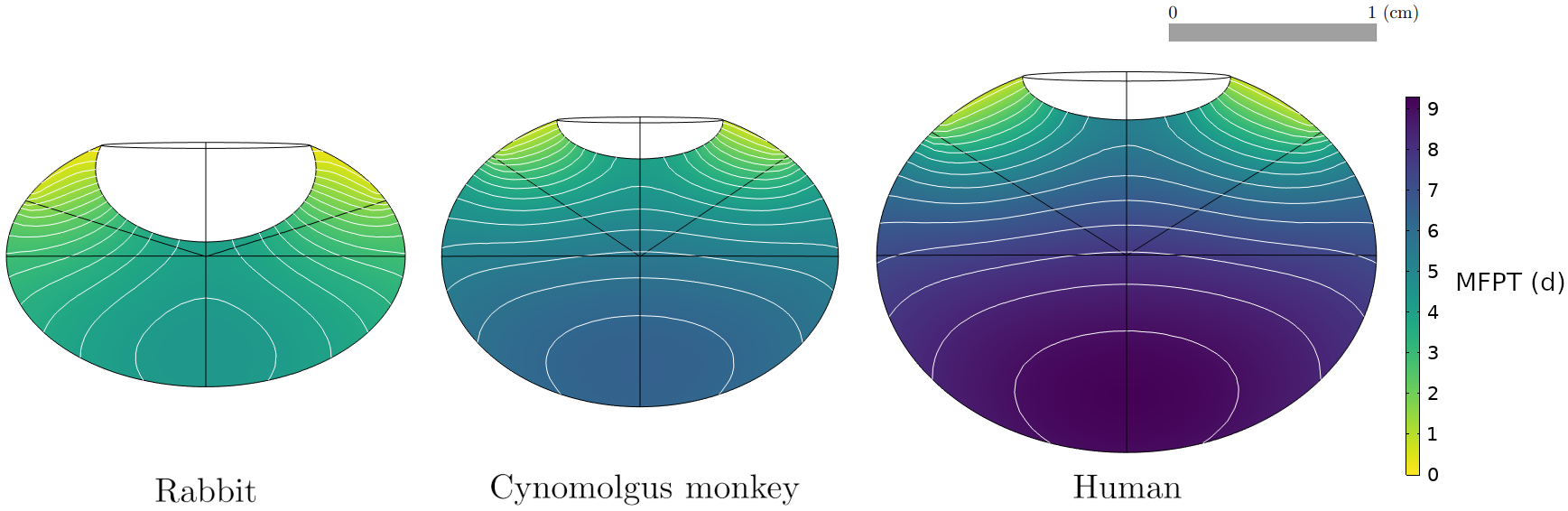}
    \end{subfigure}
    \caption{Numerical solution of the mean first passage time (MFPT) for a Fab molecule in different species, as a function of injection site, with the parameterisation of Table~\ref{tab:params_lit_review} and~\ref{tab:parameters_diff_perm_fab_igg}. Contour lines of MFPT are in white, while the colourmap indicates the MFPT value at any given point in the vitreous chamber. Black lines are associated with the construction of the model geometry (Figure~\ref{fig:geometries_uptoscale}).}
    \label{fig:mfpt_species}
\end{figure}

\begin{figure}
    \centering
    \begin{subfigure}{\linewidth}
        \centering
        \includegraphics[width=\linewidth]{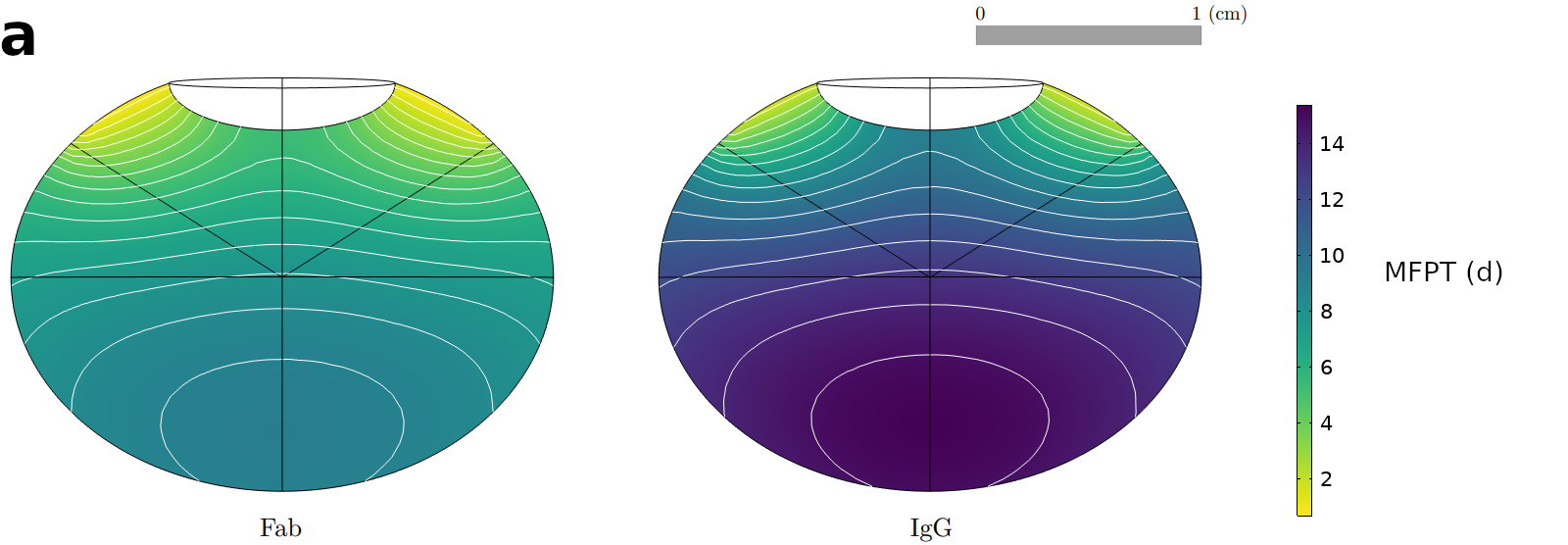}
    \end{subfigure}
    \\
    \begin{subfigure}{\linewidth}
        \centering
        \includegraphics[width=\linewidth]{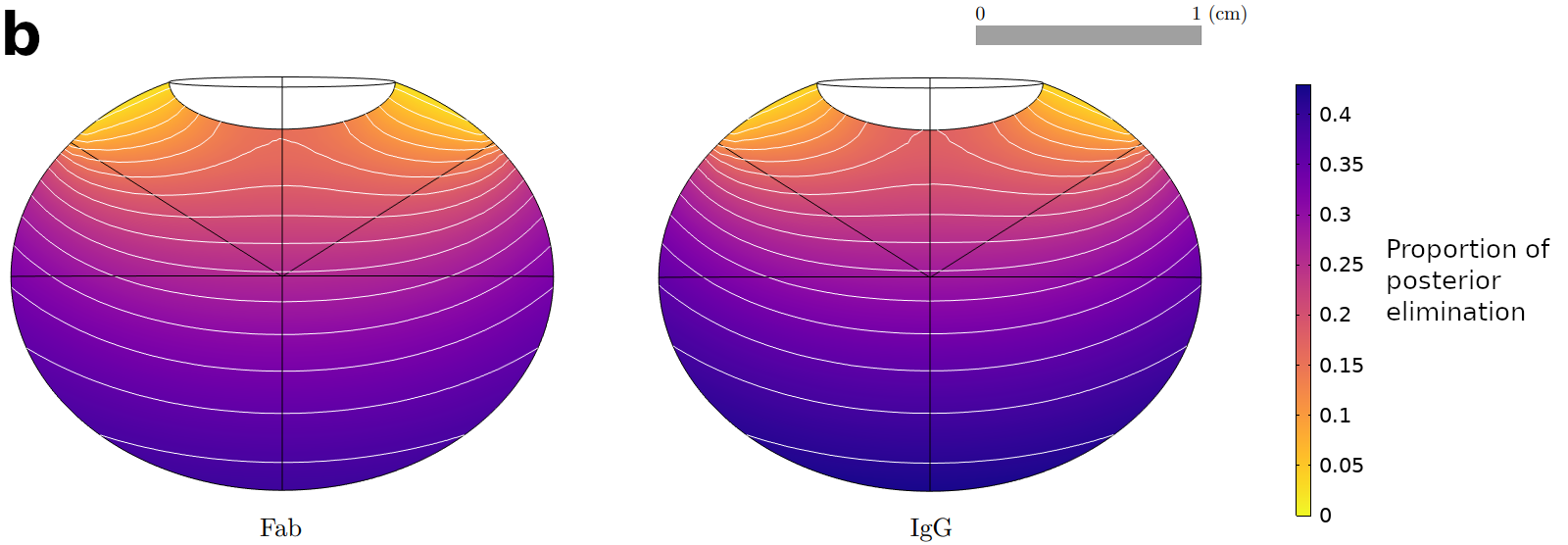}
    \end{subfigure}
    \caption{Numerical solution of the MFPT for Fab and IgG molecules in the human emmetropic eye model (\textbf{a}), and of the proportion of the drug dose exiting through the vitreous-retina interface (\textbf{b}), as a function of injection site, using the parameterisation of Table~\ref{tab:params_lit_review} and~\ref{tab:parameters_diff_perm_fab_igg}.}
    \label{fig:human_fab_igg_mfpt_prop}
\end{figure}

\begin{figure}
    \centering
    \begin{subfigure}{\linewidth}
        \centering
        \includegraphics[width=\linewidth]{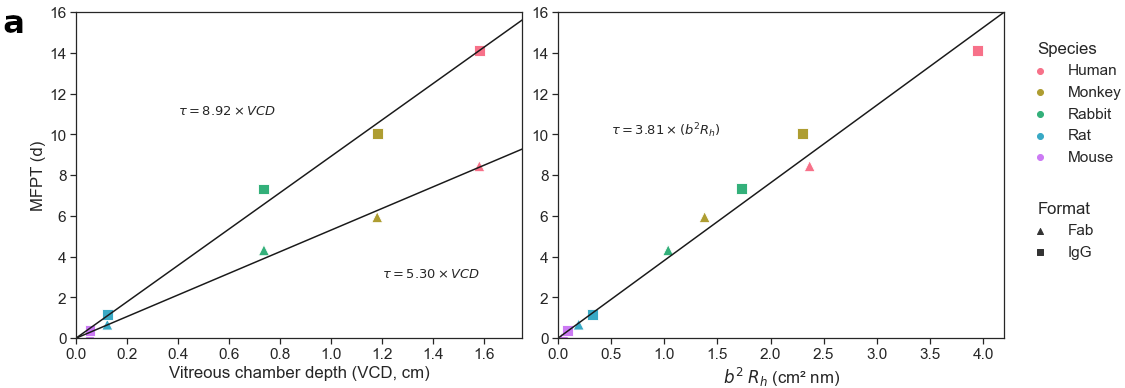}
    \end{subfigure}
    \\
    \vspace{1cm}
    \begin{subfigure}{\linewidth}
        \includegraphics[width=0.98\linewidth]{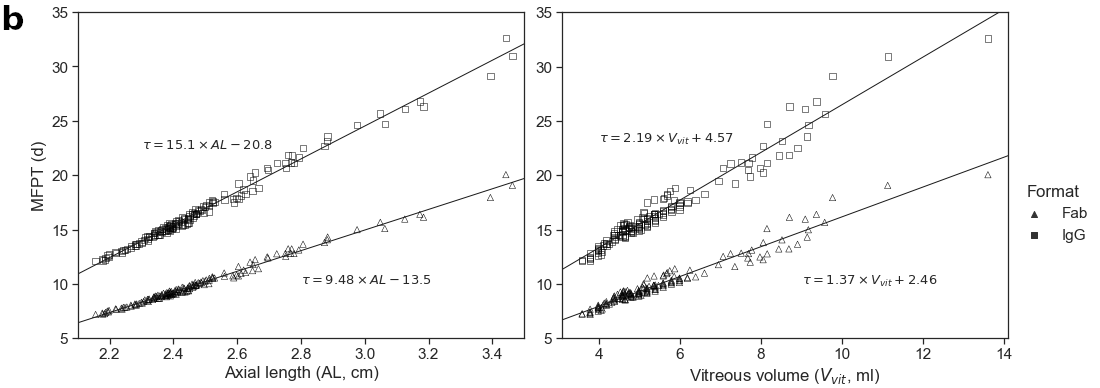}
    \end{subfigure}
    \caption{Numerical solution (symbols) and linear regressions (lines) of the MFPT for different molecular formats injected at $P_m$, using the parameterisation of Table~\ref{tab:params_lit_review} and~\ref{tab:parameters_diff_perm_fab_igg}. The MFPT is shown for (\textbf{a}) the species-specific geometries of Figure~\ref{fig:geometries_uptoscale}, and (\textbf{b}) the ensemble of human eye models of Figure~\ref{fig:set_human_eyes_data}.}
    \label{fig:mfpt_against_b2_Rh_species_humans}
\end{figure}

To obtain numerical solutions for the MFPT, we solved equations~\eqref{eq:mfpt_pde} with parameter values defined in Table~\ref{tab:params_lit_review} and~\ref{tab:parameters_diff_perm_fab_igg}, using the ocular geometries of each species (Figure~\ref{fig:geometries_uptoscale}). The results for the Fab molecule are illustrated in Figure~\ref{fig:mfpt_species}. In each model, the MFPT is maximised for an injection site at the back of the vitreous, and decreases for injections closer to the aqueous humour. The same behaviour was observed for the MFPT of an IgG molecular format, with overall longer residence times. Figure~\ref{fig:human_fab_igg_mfpt_prop}\textbf{a} compares the MFPT in the human model for a Fab and IgG molecule, with the maximum MFPT being 9 days and 14 days, respectively.
\\
\\
We observed that the MFPT decreases with eye size, with a MFPT of less than one day for all injection sites in the mouse and rat models (Figure~\ref{fig:mfpt_species}). The global sensitivity analysis identified that, within the uncertainty range of each parameter, the length of the semi-axis b was the most influential for the MFPT for an injection at $P_m$ (results shown in SI.C).\\
\\
To refine the comparison of the MFPT across species, the MFPT for a Fab and IgG molecular format was solved for an injection at $P_m$ and was plotted against the vitreous chamber depth measure of each species (Figure~\ref{fig:mfpt_against_b2_Rh_species_humans}\textbf{a}). The corresponding linear regressions were derived and constrained to go through the origin, as the intercept confidence interval included the origin, and as it is the expected theoretical behaviour. Following the analysis in~\cite{Caruso2020Ocular}, the MFPT was also plotted against the square value of the semi-axis~$b$ multiplied by the hydrodynamic radius ($b^2 \times × R_h$) (Figure~\ref{fig:mfpt_against_b2_Rh_species_humans}\textbf{a}), and a linear relationship for the MFPT across species and molecular formats was obtained, with a regression going through the origin and a slope of 3.81~days/(cm$^2$~nm). The equation~\eqref{eq:link_mfpt_t1/2} was used to obtain a slope of 2.64~days/(cm$^2$~nm) for $t_{1/2}$.
\\
\\
Considering an injection of a Fab or an IgG at $P_m$, Figure~\ref{fig:mfpt_against_b2_Rh_species_humans}\textbf{b} shows the MFPT as a function of the axial length (AL) and of vitreous volume for the ensemble of human eye models. We found that the pathological myopia eyes did not significantly affect the trends found in the ensemble of healthy human eyes, with a change of slope of less than 10\% for each molecular format when we excluded the eye geometries constructed using the pathological myopia measurements from Zhou et al.~\cite{Zhou2020Quantitative} (results presented in SI.D). Comparing the two panels of Figure~\ref{fig:mfpt_against_b2_Rh_species_humans}\textbf{b}, we see that the MFPT is better predicted by the linear relation with the AL than with the vitreous volume, with the resulting residence times being more sparsely distributed when plotted against the vitreous volume.
\\
\\
Using equation (2.2), we derived $t_{1/2}$ for all species from the MFPT for an injection located at $P_m$. The results are summarised in Table~\ref{tab:mfpt_half-life_results}, along with the experimental $t_{1/2}$ for each species. For the human eye model, we estimated a range of $t_{1/2}$ using the MFPT for hypermetropic to myopic eyes illustrated in Figure~\ref{fig:mfpt_against_b2_Rh_species_humans}\textbf{b}, hence excluding pathologically myopic eyes, and found this estimated range to be broadly consistent with the spread of $t_{1/2}$ observed in normal eyes. For the human, cynomolgus monkey and rabbit eye models, a range of $t_{1/2}$ was also estimated by solving the MFPT within the injection region (identified in Figure~\ref{fig:geometries_uptoscale}). Agreement between simulation and experiment holds for the human and the rabbit, whereas the model’s half-lives are overestimated for the cynomolgus monkey and the rat, and underestimated for the mouse.\\
\\
\begin{landscape}
\begin{table}[]
\small
\begin{spacing}{1.5}
    \centering
    \begin{tabular}{|m{2cm}  m{2.3cm} | M{1.9cm}  M{2cm} | M{4cm} | M{2cm} M{2cm} | M{4cm} |}
    \hline
        \multirow{3}{*}{Species} & & \multicolumn{3}{|c|}{Fab} & \multicolumn{3}{|c|}{IgG}  \\ \cline{3-8}
        && \multicolumn{2}{c|}{Modelling results (days)} & Experimental results (days) & \multicolumn{2}{c|}{Modelling results (days)} & Experimental results (days) \\
        && MFPT & $t_{1/2}$ & $t_{1/2}$ & MFPT & $t_{1/2}$ & $t_{1/2}$ \\ \hline
        
        Mouse & Midpoint $P_m$ & 0.20 & 0.14 & 0.86 \cite{Bussing2023Pharmacokinetics} & 0.36 & 0.25 & NA \\ \hline
        
        Rat & Midpoint $P_m$ & 0.67 & 0.46 & NA & 1.18 & 0.82 & 0.341 \cite{Chuang2010Serum, Hutton2016Mechanistic}\\ \hline

        \multirow{2}{*}{Rabbit} & Midpoint $P_m$ & 4.32 & 2.99 & \multirow{2}{*}{3.0 (2.75, 3.31) \cite{Caruso2020Ocular}} & 7.32 & 5.07 & \multirow{2}{*}{5.4 (4.17, 7.06) \cite{Caruso2020Ocular}} \\ \cline{3-4} \cline{6-7} 
        & Injection range & (4.16, 4.39) & (2.88, 3.04) & & (7.08, 7.43) & (4.91, 5.15) & \\ \hline

        Cynomolgus & Midpoint $P_m$ & 5.94 & 4.12 & \multirow{2}{*}{2.4 (2.17, 2.9) \cite{Caruso2020Ocular}} & 10.03 & 6.95 & \multirow{2}{*}{3.3 (2.8, 3.90)} \cite{Caruso2020Ocular} \\ \cline{3-4} \cline{6-7}
        monkey & Injection range & (4.94, 6.42) & (3.42, 4.45) & & (8.43, 10.76) & (5.84, 7.46) & \\ \hline
        
        Human & Midpoint $P_m$ & 8.44 & 5.85 & \multirow{3}{*}{6.5 (5.24, 8.6) \cite{Caruso2020Ocular}} & 14.12 & 9.79 & \multirow{3}{*}{9.3 (7.16, 11.67) \cite{Caruso2020Ocular}} \\ \cline{3-4} \cline{6-7}
        (emmetropic) & Injection range & (6.76, 9.22) & (4.69, 6.39) & & (11.45, 15.31) & (7.94, 10.61) & \\ \cline{3-4} \cline{6-7}
        & Ensemble range & (7.52, 10.74) & (5.21, 7.44) & & (12.63, 17.79) & (8.75, 12.33) &  \\ 
        \hline

    \end{tabular}
    \caption{MFPT and estimated $t_{1/2}$ (using equation (2.2)) for an injection at midpoint $P_m$, for the injection range (injection location region identified in Figure~\ref{fig:geometries_uptoscale}) reporting the min-max results, and for the ensemble range (for an injection at $P_m$ in the ensemble of human eye models excluding the pathological myopia data) reporting the 5th-95th percentile. The experimental  $t_{1/2}$ (mean, with lower and upper bound of found interval) for each species and molecular format are also reported.}
    \label{tab:mfpt_half-life_results}
\end{spacing}
\end{table}
\end{landscape}

\subsection*{Drug elimination}
\begin{figure}
    \centering
    \begin{subfigure}{\linewidth}
        \centering
        \includegraphics[width=\linewidth]{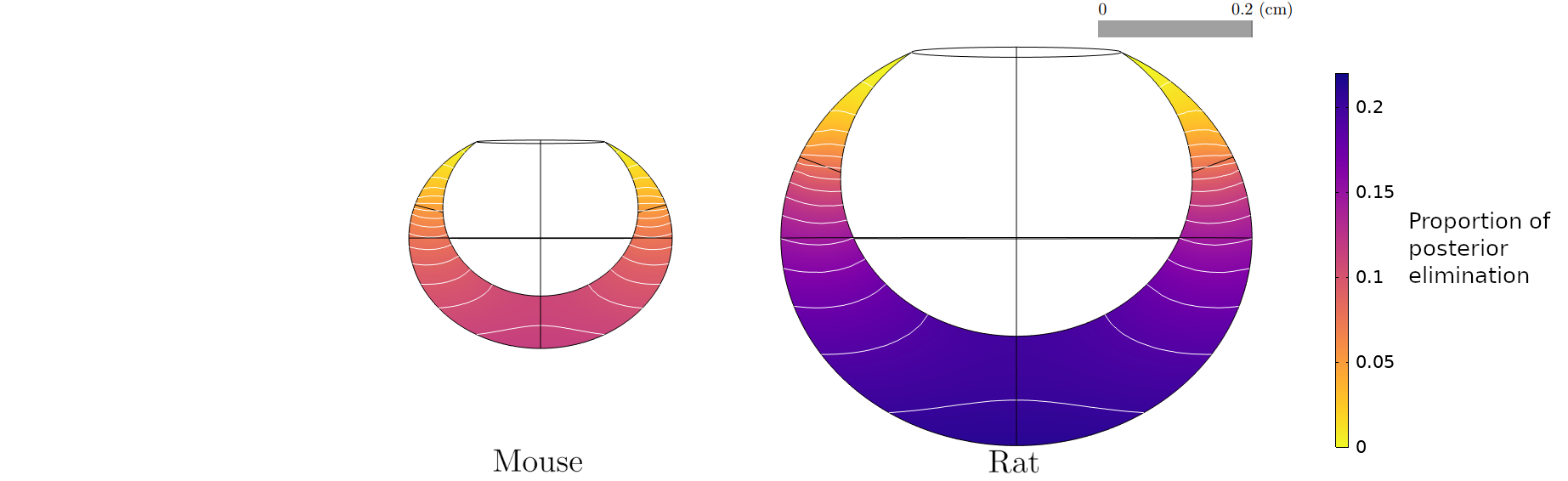}
    \end{subfigure}
    \\
    \vspace{1cm}
    \begin{subfigure}{\linewidth}
        \centering
        \includegraphics[width=\linewidth]{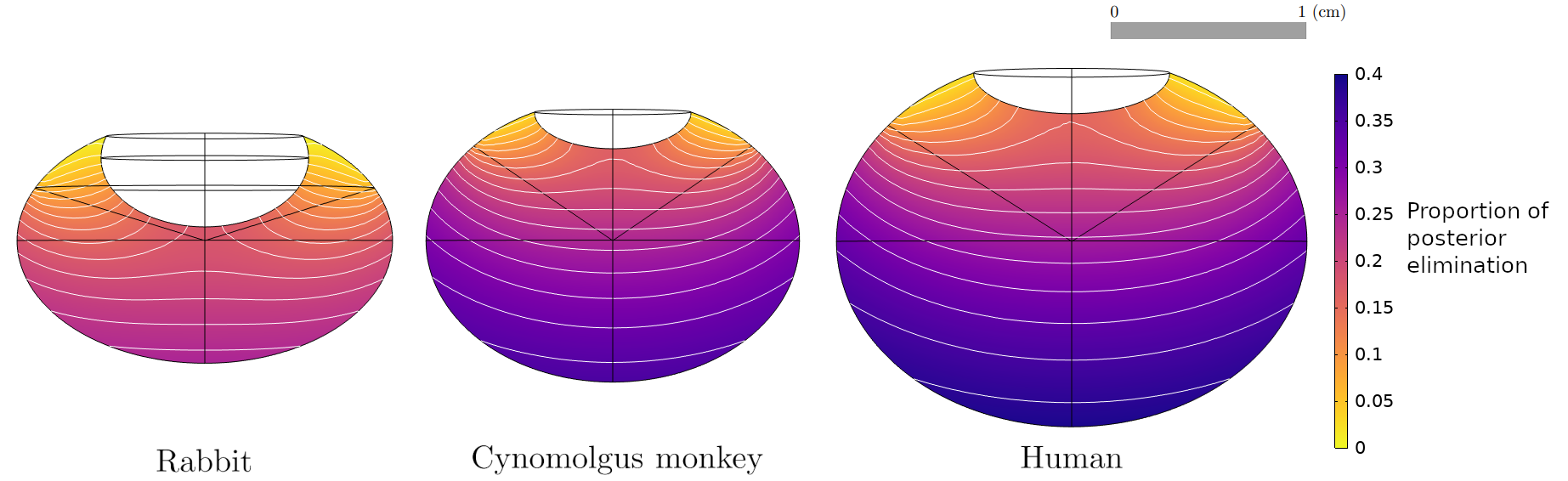}
    \end{subfigure}
    \caption{Numerical solution and contour lines of the proportion of drug dose exiting through the vitreous-retina interface (unitless), for a Fab molecule and for each species, as a function of injection site, with the parameterisation of Table~\ref{tab:params_lit_review} and~\ref{tab:parameters_diff_perm_fab_igg}, and the geometries in Figure~\ref{fig:geometries_uptoscale}.}
    \label{fig:distribution_exit_retina_species}
\end{figure}

\begin{figure}
    \centering
    \includegraphics[width=\linewidth]{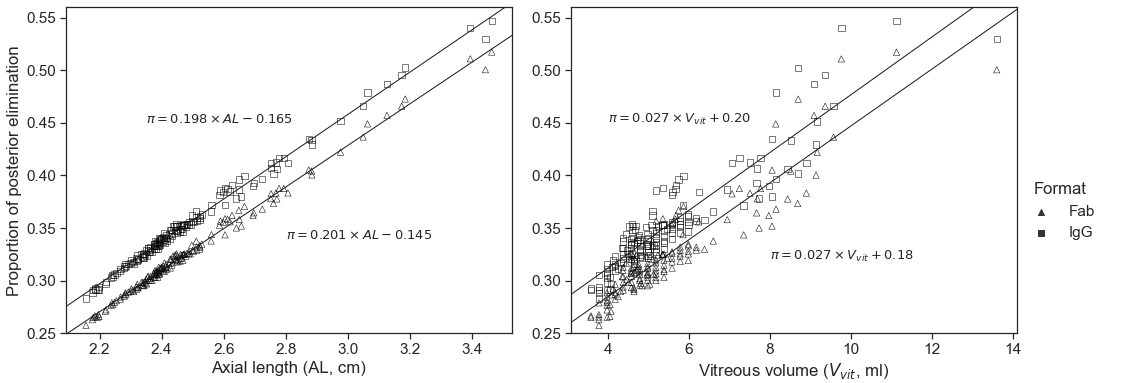}
    \caption{Numerical solution and regression lines of the proportion of drug dose exiting through the vitreous-retina interface in the ensemble of human eye models (as identified in Figure~\ref{fig:set_human_eyes_data}) for an injection at $P_m$ and with parameters of Table~\ref{tab:parameters_diff_perm_fab_igg}, as a function of the AL and vitreous volume.}
    \label{fig:distribution_exit_retina_human_eyes}
\end{figure}

To obtain the numerical solutions for the proportion of drug exiting through the vitreous-retina interface, equations (2.3) were solved with the parameters and ocular geometries of Table~\ref{tab:parameters_diff_perm_fab_igg} and Figure~\ref{fig:geometries_uptoscale}, respectively. The results for the Fab molecule are illustrated in Figure~\ref{fig:distribution_exit_retina_species}. The maximum contribution of posterior elimination varied across species, with up to 40\% of dose permeating the vitreous-retina interface in the human, versus 12\% in the mouse, while the posterior elimination from the injection point $P_m$ varied from 10\% to 30\% across species. Injection sites located at the back of the eye were associated with higher proportions of posterior elimination, which decreased with the distance from the anterior segment. Similar results were obtained for IgG, as visible in Figure~\ref{fig:human_fab_igg_mfpt_prop}\textbf{b} for the emmetropic human eye. 
\\
\\
To study the influence of inter-individual anatomical differences, the ensemble of human eye models was solved for the posterior elimination after injection at $P_m$ (Figure~\ref{fig:distribution_exit_retina_human_eyes}). The posterior contribution to ocular elimination showed a strong correlation with the axial length (AL) and vitreous volume for both Fab and IgG, with poor separation between molecular formats for the latter.
\\
\\

\subsection*{Conditional MFPT}
\begin{figure}
    \centering
    \includegraphics[width=0.85\linewidth]{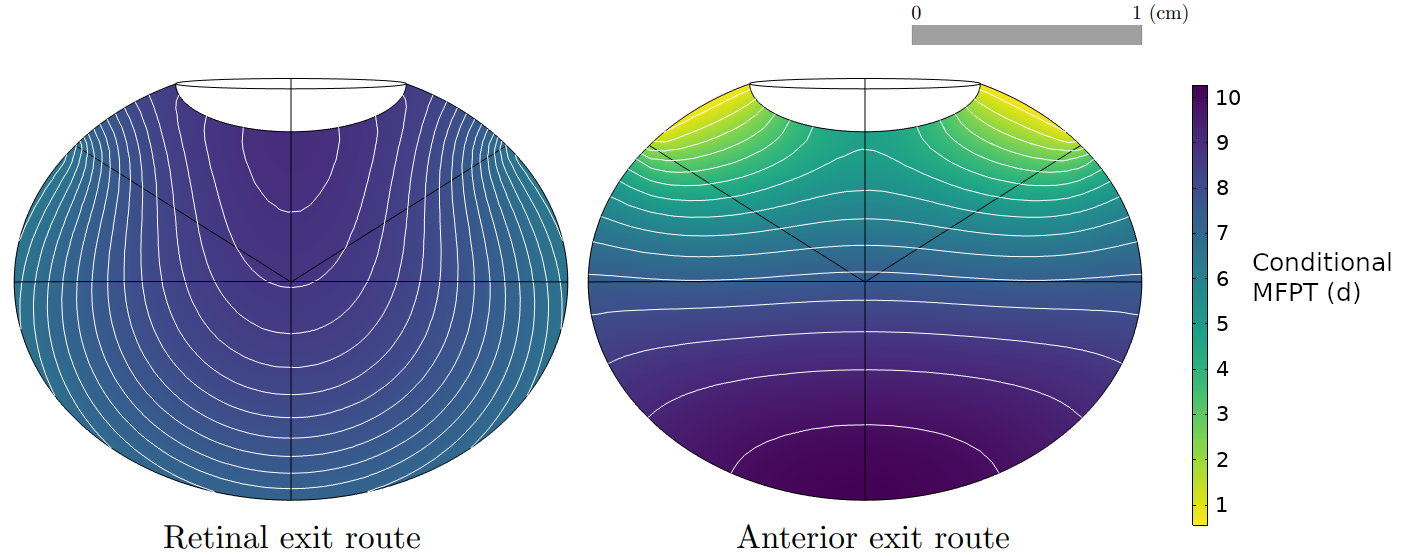}
    \caption{Numerical solution and contour lines for the MFPT, conditional on exiting through the vitreous-retina (left) and vitreous-aqueous humour (right) interfaces, in the human eye model for a Fab molecule, as a function of injection site, with the parameterisation of Table~\ref{tab:params_lit_review} and~\ref{tab:parameters_diff_perm_fab_igg}.}
    \label{fig:cond_mfpt_human_macula_retina_aqueous}
\end{figure}

To obtain numerical solutions for the conditional MFPT, which gives the duration time conditioned on the exit rate, equations (2.3) and (2.4) were solved with parameter values given in Table~\ref{tab:parameters_diff_perm_fab_igg} for a Fab molecule, using the human eye geometry. Figure~\ref{fig:cond_mfpt_human_macula_retina_aqueous} shows the results for the conditional MFPT. The conditional MFPT for the drug exiting through the vitreous-retina interface has a lower variation range, with exit times varying between 6.5 and 9 days, and has very different contour plots to those of the unconditional MFPT (Figure~\ref{fig:mfpt_species},~\ref{fig:human_fab_igg_mfpt_prop}). In contrast, the conditional MFPT for the drug exiting through the vitreous-aqueous humour interface has a similar range of values and contour plots compared with the unconditional MFPT, indicating that the dynamics in the MFPT solutions could be dominated by the dynamics of the anterior elimination pathway. This behaviour was observed in all modelled species (results presented in SI.D), and held for the IgG molecular format.\\
\\

\section*{Discussion}
We have given the governing equations for the mean first passage time, MFPT, the MFPT conditioned on the exit route, and the proportion of drug exiting through subsections of the eye, for Fab and IgG molecular formats. We built realistic 3D eye geometries based on ocular measurements, and confirmed their anatomical accuracy by comparing them to MRI images. We solved the equations on the ocular geometries, and we analysed our results to assess the potential influence of spatial parameters on ocular drug residence times. We linked this analysis to ocular half-life and contrasted it with experimental data from nonclinical species and humans.
\\
\\
This study’s aims have included exploring the importance of injection site location and individual variation for IVT administration of protein therapeutics. The model simulations show that the injection location had a significant effect on the MFPT in all species (Figure~\ref{fig:mfpt_species}). This is in line with a previous computational study, where the influence of four different injection locations in a human eye model was investigated~\cite{Friedrich1997Drug, Friedmann2023Models}, concluding that injections at the back of the eye induced higher drug concentrations in the first 24 hours after injection~\cite{Friedrich1997Drug} and maximal drug absorption at the macula~\cite{Friedmann2023Models}. In the rodent models, the MFPT was higher for injection sites centred on the vitreous chamber depth along the optical axis behind the lens in the vitreous, however, these may not be accessible due to the curvature of the lens. We postulate that variations in reported ocular half-lives (both in clinical and nonclinical species) could be partially explained by different injection site locations in the vitreous chamber. In the human eye, the results indicate a large posterior region of the vitreous body where IVT injections are expected to yield maximal drug residence time and retinal permeation (Figure~\ref{fig:mfpt_species},~\ref{fig:human_fab_igg_mfpt_prop},~\ref{fig:distribution_exit_retina_species}). Conversely, injections closer to the anterior segment lead to lower elimination across the retina. It is expected that alternative ocular delivery approaches to IVT injection, such as implants and sustained release formulations, similarly ought to target the central and posterior vitreous for improved retinal exposure.
\\
\\
The results of this work agree with previous reports that IVT PK depends on ocular and molecular size \cite{Caruso2020Ocular, Crowell2019Influence}. A linear relation was found between the MFPT and the vitreous chamber depth, for both Fab and IgG molecular formats (Figure~\ref{fig:mfpt_against_b2_Rh_species_humans}\textbf{a}). Also, the $t_{1/2}$ was linearly correlated to the product of the eye semi-axis $b$ squared and the protein $R_h$, with the associated regression line having a slope of 2.64 days/(cm$^2$~nm). This is comparable to the range previously reported by Caruso et al.\cite{Caruso2020Ocular} (1.3-2.4~days/(cm$^2$~nm)) for the linear regression of half-life on $r_{vit}^2 \times R_h$ in different species, where $r_{vit}$ is the vitreal radius. The experimental estimates of that study suggest the relationship is species-specific, with the minipig exhibiting the steepest slope (2.4 days/(cm$^2$~nm)) and the rabbit, human, rat and monkey displaying shallower lines (2.1, 1.8, 1.6 and 1.3 days/(cm$^2$~nm), respectively). Interestingly, this sequence of species is not ordered by vitreous size, for example, the vitreal volume in the rabbit is smaller (and thus, diffusion distance is shorter) than both human and monkey vitreous (Table~\ref{tab:params_lit_review}). This suggests that factors other than eye size must play a role in determining the PK differences observed across species. Vitreal chamber shape and eccentricity have been proposed as possible determinants \cite{Caruso2020Ocular}, a notion that is not supported by the present results. In fact, in this work the anatomically realistic description of ocular geometry produced collinear estimates of residence time across species (Figure~\ref{fig:mfpt_against_b2_Rh_species_humans}\textbf{a}). Also, the model simulations yielded half-life values aligned with inter-species differences in vitreal volume, i.e. longer half-lives in larger eyes (Table~\ref{tab:mfpt_half-life_results}). While the human and rabbit $t_{1/2}$ estimates are close to the experimental values, a discrepancy is found for the cynomolgus monkey and for the rodents, although the $t_{1/2}$ in the latter is not as well established as in the larger species. We conjecture that the partial mismatch between experimental data and simulation outcomes can be attributed to the assumptions made regarding the permeability parameters. Notably, including species-specific permeability parameters in the present model can significantly influence the relative contribution of the anterior and posterior pathway to drug elimination. The proportion of drug exiting through the posterior pathway is directly linked to the two permeability parameters within the model. The relative contribution in exit pathways has been previously proposed as a potential determinant of $t_{1/2}$ in ocular PK \cite{Caruso2020Ocular, Crowell2019Influence}. Moreover, insights from a prior study on topically applied small molecules indicate variations in ocular tissue permeability across different species \cite{Loch2012Determination}. Extending this understanding to IVT macromolecules, species-specific permeability data may be necessary to obtain more accurate modelling results. While a few studies have previously investigated the effect of molecular size on the permeability of ocular tissues \cite{Yoshihara2017Permeability, Kim2021Permeability, Ramsay2019Role}, there is a lack of comparative studies on the permeability of IVT macromolecules across species.
\\
\\
The global sensitivity analysis showed that the axial length is the most influential parameter on residence time across species, confirming the importance of eccentricity in modelling the vitreous chamber. On the other hand, the spherical approximation used in several previous works~\cite{Hutton2016Mechanistic,Jooybar2014Computational,Dosmar2021Compartmental,Khoobyar2021Analytical,Li2022Drug} implies that the semi-axis $b$ has the same length as semi-axis $a$. By way of example, if we compare the injection of a Fab molecule at $P_m$ in the human emmetropic eye model ($a=1.1275$ cm, $b=0.889$ cm) and in a spherical model of same vitreous volume ($a=b=1.043$ cm), we find a meaningful discrepancy in the estimate of the half-life (respectively $5.85$ and $7.16$ days). Hence, we recommend that future models, aiming to explore PK across various species, do away with the spherical approximation.\\
\\
Previous work has investigated the impact of eye size on PK in diverse animal species. To the best of our knowledge, this is the first report showing a significant impact of eye shape and eccentricity within the same species, namely humans. Ocular half-life estimates are known to exhibit sizable differences between patients in clinical studies. For example, Avery et al.~\cite{Avery2017Systemic} reported a mean $t_{1/2}$ and standard deviation of 5.8 (4.0,~7.6) days for the Fab ranibizumab, and Meyer et al.\cite{Meyer2011Intraocular} reported an average and 95\% confidence interval of 11.17 (8.7, 18.2) days following a 3 mg injection of the IgG bevacizumab. On the other hand, a previous study  in 41 eyes found no correlation between intraocular drug concentration and axial length (AL)\cite{Krohne2015Influence}. Such contrast of results motivated this research to better understand whether inter-individual differences in ocular geometry may impact PK. In the ensemble of human eyes having different volumes and axial lengths, our model showed a large variation in residence time (Figure~\ref{fig:mfpt_against_b2_Rh_species_humans}\textbf{b}), contributing to explain the experimental inter-individual variability. The different slopes between the Fab and IgG molecules suggest that the variability in the AL of the eye elongation is more influential on the residence time for molecules with slower diffusion. The linear relation found between MFPT and AL suggests that measuring the AL is sufficient to obtain an estimate of the residence time for any given human eye. This has promising implications for clinical practice, as AL measurements can be obtained in patients with relative ease, more so than for vitreous volume. Optical biometry, both reasonably simple and cost-effective, could serve as a potential stride towards personalised treatment by furnishing insights into individual durability of ocular exposure and pharmacology.
\\
\\
The model also suggests that the proportion of posterior elimination varies greatly between species (Figure~\ref{fig:distribution_exit_retina_species}), which appears to be mostly correlated to the distance between the injection site and the vitreous-aqueous humour interface. In the rabbit eye model, the posterior pathway contribution to drug elimination was 19\% to 23\% (Figure~\ref{fig:distribution_exit_retina_species}), for an injection within the region identified in Figure~\ref{fig:geometries_uptoscale}, which is in line with previous experimental estimates. A prior study in rabbit eyes estimated the posterior clearance as 3\% to 20\% of the dose administered by IVT injection \cite{delAmo2015Rabbit}. Another computational model calculated the percentage of Fab molecules exiting through the RPE to be 12.7\% of the IVT dose in a rabbit experiment \cite{Hutton2017Ocular}. To our knowledge, current estimates are informed by experimental data obtained in rabbits, while Figure~\ref{fig:distribution_exit_retina_species} highlights clear species differences, making it necessary to further study the elimination pathways in other species. In the ensemble of human eyes, we found that posterior elimination of both Fab and IgG formats is linearly correlated with the axial length and vitreous volume (Figure~\ref{fig:distribution_exit_retina_human_eyes}), strengthening the notion that individual variations in eye shape may influence drug disposition and pharmacology.
\\
\\
The results of the MFPT conditioned on the exit route have provided further information on the dynamics of the MFPT. In all species, the solutions of the conditional MFPT show that the clearance pathway through the vitreous-aqueous humour interface is dominating the behaviour in the MFPT solutions. Furthermore, the model suggests that drug molecules leaving through the retina are spending more time in the vitreous chamber than molecules exiting through the aqueous humour (Figure~\ref{fig:cond_mfpt_human_macula_retina_aqueous}), and thus the duration time of drug that exits into the target region is longer than the mean duration time. Hence, the half-life underestimates the duration of drug in the vitreous that reaches the target.
\\
\\
In conclusion, the residence time and the posterior elimination were studied across multiple species used in ocular research and drug development, with the aim to strengthen the inter-species translation of pharmacokinetic and pharmacodynamic studies. The anterior pathway was identified as the predominant route of drug elimination, and the contribution of the posterior pathway varied significantly across species. The injection location was found to be highly influential in the drug kinetics, and maximum efficacy was obtained for injections in the posterior vitreous. Additionally, we showed that the variability in vitreous chamber size and shape in human eyes can lead to significant differences in drug residence times and proportion of posterior elimination. The methodology developed in this study emerges as a potent framework for characterising the vitreal transport dynamics of current ocular therapeutics. By combining our methodology with species-specific measurements of posterior permeabilities, it would be possible to investigate the efficacy of emerging ocular therapeutics.\\
\\

\section*{Acknowledgments}
The authors acknowledge the support of the Natural Sciences and Engineering Research Council of Canada (NSERC) [567922], of the Engineering and Physical Sciences Research Council (EPSRC) [EP/S024093/1], and of Roche Pharma Research and Early Development. The authors would like to thank Nora Denk for insightful conversations around intravitreal injection procedures. They also thank Norman Mazer for his careful review and input to this manuscript.

\newpage
\begin{singlespace}
\bibliography{paper_biblio}
\end{singlespace}

\end{document}


\maketitle
\doublespacing

\appendix
\counterwithin{figure}{section}
\counterwithin{table}{section}

\section{Derivation of the relation between the mean first passage time and the ocular half-life}
\begin{figure}
    \centering
    \includegraphics[width=0.6\linewidth]{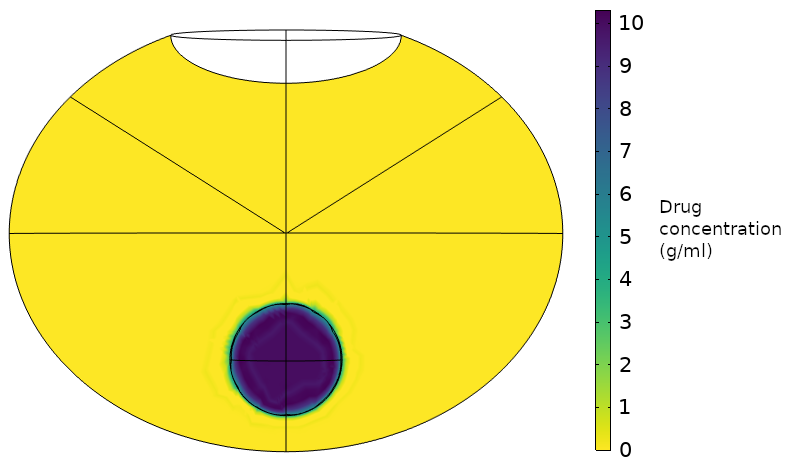}
        \caption{Initial condition for the diffusion simulation for an injection at the back of the vitreous in the human eye. The parameters used to produce this plot are in Table 1 and Table 3, with the geometry of the human eye illustrated in Figure 2.}
        \label{fig:SI_diffusion_initial_condition}
\end{figure}

\begin{figure}
    \begin{subfigure}{0.49\linewidth}
    \centering
    \includegraphics[width=\linewidth]{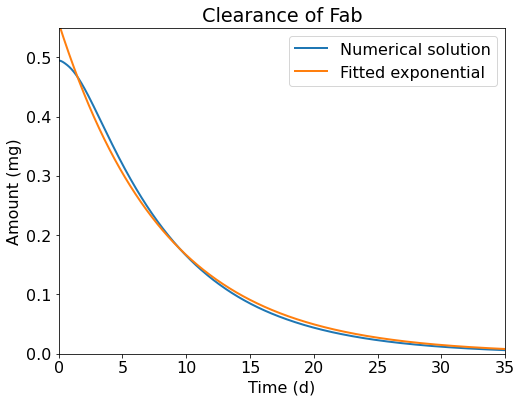}
    \end{subfigure}
    \begin{subfigure}{0.49\linewidth}
    \centering
    \includegraphics[width=\linewidth]{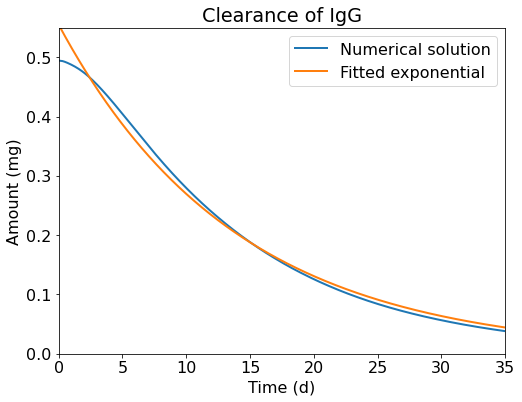}
    \end{subfigure}
    \caption{Numerical solutions of the diffusion simulations for an injection at the back of the eye in the human eye, for a Fab (left) and an IgG (right) molecule, with the fitted exponential decay function used to calculate directly the ocular half-life.}
    \label{fig:SI_diffusion_sol_fab_igg}
\end{figure}

The half-life, $t_{1/2}$, is the time required for a quantity that is exponentially decaying to fall to one half of its initial value. In the context of this application, the ocular $t_{1/2}$ characterises the drug's rate of clearance. In experimental studies, $t_{1/2}$ is usually calculated using the coefficients of the exponential curve fitted to the collected concentration data, using concentration in the aqueous as a proxy for concentration in the vitreous~\cite{Caruso2020Ocular}. Here, we derive an equation to link the mean first passage time (MFPT) with $t_{1/2}$.\\
\\
Let $c(t)$ be the total quantity of drug inside the vitreous at time $t$, for a specified injection site $\boldsymbol{x_0}$, and $c_0$, the initial concentration of drug. The proportion of drug remaining in the eye at time $t$ is $\frac{c(t)}{c_0}$. Let $T$, a random variable, be the first passage time for injection at location $\boldsymbol{x_0}$. Treating all drug molecules as equivalent (so considering the proportion of drug exiting instead of the probability of one particle exiting),  we have
\begin{align*}
    &\text{ Prob}(T > t) = \text{Proportion of drug remaining at time } t  = \frac{c(t)}{c_0}.
\end{align*}
Thus
\begin{align*}
    \text{ Prob}(T < t) = 1 - \frac{c(t)}{c_0},
\end{align*}
and hence
\begin{align*}
    &\text{ Prob}(T \in [t, t+ \delta t]) = \text{Prob}((T < t + \delta t) \cap (T \nless t)) \\
    & \hspace{3.15cm} = \text{Prob}(T < t + \delta t) - \text{Prob}(T < t) \\
    & \hspace{3.15cm} = - \frac{1}{c_0} \left( c(t+\delta t) - c(t) \right) \\
    & \hspace{3.15cm} = - \frac{1}{c_0} \frac{dc(t)}{dt} \delta t + \mathcal{O}(\delta t^2),
\end{align*}
where the last line was obtained using a Taylor approximation around $t$ and $\delta t$ is a small time increment. Therefore, by taking the limit $\delta t \rightarrow 0$, the probability density function for the first passage time $T$ is
\begin{equation*}
    f_T (t) = - \frac{1}{c_0} \frac{dc(t)}{dt},
\end{equation*}
for the previously specified initial injection location $\boldsymbol{x_0}$. By definition of the mean first passage time $\tau$ as being the expected value of the first passage time, we have
\begin{align}
    \tau = \int_0^\infty t \left( - \frac{1}{c_0} \frac{dc}{dt} \right) dt = \frac{1}{c_0} \int_0^\infty c \, dt,
    \label{eq:mfpt_total_quantity_drug_relation}
\end{align}
assuming $c \rightarrow 0$ faster than $1/t$ as $t \rightarrow \infty$, which is justified as we are expecting a behaviour similar to an exponential decay for $c(t)$.\\
\\
For $c(t)$ decreasing exponentially, with initial concentration $c_0$, the drug concentration can be expressed as
\begin{equation}
    c(t) = c_0 \, \mathrm{e}^{- \lambda t},
    \label{eq:c_exponential_decay}
\end{equation}
where $\lambda$ is the decay rate (or, more accurately, the clearance rate of the drug from the vitreous humour). The corresponding half-life is
\begin{equation}
    t_{1/2} = \frac{\ln{2}}{\lambda}.
    \label{eq:half-life}
\end{equation}
\\
\\
Using equations~\eqref{eq:c_exponential_decay} and~\eqref{eq:half-life} in equation~\eqref{eq:mfpt_total_quantity_drug_relation}, we obtain
\begin{equation*}
    \tau = \frac{1}{c_0} \int_0^\infty c_0 \, \mathrm{e}^{- \lambda t} dt = \frac{1}{\lambda} = \frac{t_{1/2}}{\ln{2}}.
\end{equation*}
Hence, for a concentration decreasing exponentially at all time, the relation between the MFPT and the ocular half-life is
\begin{equation}
    t_{1/2}(\boldsymbol{x_0}) = (\ln{2}) \tau(\boldsymbol{x_0}).
    \label{eq:half-life_mfpt}
\end{equation}
where $\boldsymbol{x_0}$ is the injection location.\\
\\
To obtain equation~\eqref{eq:half-life_mfpt}, we made the assumption that $c(t)$ was decreasing exponentially at all time. To support the justification of this assumption, we have solved the diffusion equation for an injection of 0.5 mg of drug in 50 \textmu l liquid~\cite{Niwa2015Ranibizumab, Muether2013Long}, centered on the optical axis at the back of the vitreous, in the human eye model, for a Fab and an IgG molecule format (see Figure~\ref{fig:SI_diffusion_initial_condition}). The solutions are illustrated in Figure~\ref{fig:SI_diffusion_sol_fab_igg}, where the quantity of injected drug varies with time due to the drug clearance from the vitreous. We fitted an exponential decay function and obtained the decay rate to directly measure the ocular half-life associated with this setting. The results are summarised in Table~\ref{tab:SI_diffusion_sol_validation_half-life}.\\
\begin{table}[]
    \centering
    \begin{tabular}{| m{12cm} | c | c |}
    \hline
         Measure & Fab & IgG \\ \hline
         MFPT (days) & 9.216 & 15.308 \\
         $t_{1/2}$ estimated using the MFPT (days) & 6.39 & 10.61 \\
         $t_{1/2}$ estimated from the diffusion simulation and fitted exponential (days) & 5.73 & 9.61 \\ 
         Relative error (\%) & 11.5 & 10.4 \\ \hline
    \end{tabular}
    \caption{\raggedright Relevant quantities for the validation of equation~\eqref{eq:half-life_mfpt}.}
    \label{tab:SI_diffusion_sol_validation_half-life}
\end{table}
\\
In Table~\ref{tab:SI_diffusion_sol_validation_half-life}, the $t_{1/2}$ estimated with the MFPT (using equation~\eqref{eq:half-life_mfpt}) was obtained assuming that the quantity of drug leaving the vitreous followed an exponential decrease, whereas the second $t_{1/2}$ was obtained by fitting an exponential function to the decrease of drug quantity over time. In experimental settings, where the quantity or concentration of drug is measured over time, the half-life is obtained by the second method, i.e. by fitting an exponential function and extracting its decay rate. Hence, we considered the $t_{1/2}$ derived by the diffusion simulation to be more representative of the experimentally measured $t_{1/2}$. For an injection site at the back of the eye (which provided the largest discrepancy), we obtained differences of 10.4\% and 11.5\% between the two measures, for an IgG and a Fab molecule respectively. Considering the high uncertainty on the permeability parameters, obtained from rabbit data, we did not expect our model to have the ability of estimating the ocular half-lives with a great precision and consider a 10\% relative error introduced by our modelling framework to be acceptable.\\
\\

\section{Details on geometry construction}

\subsection{Details on geometry construction for each species} \label{sec:SI_details_geom_species}
Below is a detailed description of the construction of the canonical eye model for each species, as illustrated in Figure 2 of the main text, with parameters specified in Table 3.\\

\subsubsection*{Human}
Given experimental data for of eye geometries as a function of age, we chose to consider measures for the range of 50-95 years old, to reflect the age range of the majority of people affected by wet AMD. Using measures from the literature, canonical parameters are given as follows:
\begin{itemize}
    \item The vitreous chamber diameter was set to $2.255$ cm \cite{Atchison2004Eye}, taking the average height and width measures from Table~1 in~\cite{Atchison2004Eye} for emmetropic eyes, which yields a semi-axis of $a = 1.1275$ cm for the ellipsoid representing the vitreous chamber.
    \item The lens thickness was set to 0.3909 cm, using a linear fit for 50 year-olds from MRI measures \cite{Koretz2004Scheimpflug}.
    \item For 50 year-olds, with the linear regression from~\cite{Rosen2006Vitro}, the lens diameter was estimated to be 0.939~cm.
    \item Based on in situ MRI, we set $l_p = 50$\%, i.e. we supposed that half of the lens is situated inside the vitreous chamber cavity~\cite{Atchison2004Eye}.
    \item The optical axial length denotes the length between the retina and the cornea on the optical axis and was set to 2.30~cm, the average measure for emmetropic eyes in \cite{Atchison2004Eye}. The anterior chamber depth, that is the length between the cornea and the lens, was set to 0.3276~cm \cite{Koretz2004Scheimpflug}, using the citation's linear fit for 50 year-olds from MRI measures. We defined the semi-axis $b$ as half the length on the optical axis between the centre of the lens and retina. Subtracting the anterior chamber depth and the anterior half of the lens thickness from the axial length, we obtained
    \begin{equation*}
         b = \frac{2.30 - (0.3276 + 0.3909/2)}{2} \text{ cm} = 0.889 \text{ cm}.
    \end{equation*}
    \item For the height of the vitreous-aqueous interface, we used the estimated ratio of vitreous-aqueous surface area to the total surface area of 15\% to define $h_{va} = 0.251$ cm \cite{Hutton2016Mechanistic}.
\end{itemize}
We validated these ocular dimensions by comparing the vitrous volume and the retinal surface area with measures from the literature. The canonical model's geometry had a vitreous volume of 4.595~ml, which was in the range of vitreous volumes measured for 50 to 95-year-olds~\cite{Azhdam2020Vivo}. The constructed geometry had a retinal surface area of 10.963~cm², which was within the range of retinal surface areas measured in the literature. \\
\\

\subsubsection*{Cynomolgus monkey}
\begin{itemize}
    \item The lens diameter was set to 0.75 cm, taking the mean of the second group of cynomolgus monkeys considered by Manns et al. \cite{Manns2007Optomechanical}, which included lenses from `older donors'.
    \item The lens thickness was set to 0.351~cm, taking the mean value of Choi et al. \cite{Choi2021NormativeCyno}.
    \item Based on the longitudinal section of a cynomolgus monkey eye, we set the proportion of the lens inside the vitreous chamber cavity $l_p$ to be 50\%~\cite{Short2008Safety}.
    \item The anterior chamber depth was set to 0.309~cm~\cite{Choi2021NormativeCyno}, and the optical axial length was set to 1.841~cm~\cite{Choi2021NormativeCyno}. The semi-axis $b$ for the vitreous chamber ellipsoid was obtained by subtracting the anterior chamber depth and half of the lens thickness from the optical axial length, i.e.
        \begin{equation*}
            b = \frac{1.841 - (0.309 + 0.351/2)}{2}  \text{ cm} = 0.678 \text{ cm}.
        \end{equation*}
    \item The height of the vitreous-aqueous interface was set to 0.163~cm, so that the ratio of the surface of the vitreous-aqueous humour interface to the total surface of the vitreous ellipse was approximately 13\%~\cite{Hutton2016Mechanistic}.
    \item No experimental measure of the vitreous chamber diameter of the cynomolgus monkey was found in the literature in order to parameterise $a$. We therefore used the measure of the vitreous volume from the literature to fix $a$. In order to have a vitreous volume value of $V_\text{vit} = 2.2$ cm$^3$, we set $a=0.895$ cm~\cite{Atsumi2013Comparative}. To fit the range of vitreous volumes of 2.0 to 2.3~ml (\cite{Atsumi2013Comparative}), we set the range of $a \in [0.855, 0.915]$ cm.
\end{itemize}
In contrast with the other species, we could not use the vitreous volume to validate our ocular dimensions, as we used the literature vitreous volume to define the semi-axis of the vitreous chamber width $a$. Therefore, we validated the constructed geometry by comparing the model's retinal surface area with measures from the literature. The geometry had a retinal surface area of 6.9105~cm², which was within the range of retinal surface areas reported in the literature for the rhesus monkey (no measure could be found for the cynomolgus monkey), which ranged between 5.8 and 9.2~cm², with a mean of 7.30 cm²~\cite{Wikler1990Photoreceptor}. The rhesus monkey eyes are similar to the cynomolgus monkey eyes, with a slightly larger axial length (between 1.9 cm and 2.0 cm)~\cite{Fernandes2003Ocular}.\\

\subsubsection*{Rabbit}
\begin{itemize}
    \item In contrast to the human and cynomolgus monkey eyes, the rabbit lens has more than half of its volume inside the vitreous chamber (see MRI of rabbit eyes in \cite{Sawada2002Magnetic, Tsiapa2015High}). Guided by in situ MRI, we applied a translation of the centre of the lens of $l_T/7$ towards the centre of the vitreous ellipse, to obtain a geometry that visually matched, with approximately 2/3 of the lens inside the vitreous chamber.
    \item The lens thickness was set to 0.66 cm, the mean value of \cite{Atsumi2013Comparative}. Its range was determined by the range of measures reported in \cite{Atsumi2013Comparative, Liu1998Twenty}.
    \item The lens diameter was set to 0.995 cm, the mean value of in situ measurements in \cite{Werner2006Experimental}, and its range to the standard deviation reported in the paper.
    \item The anterior chamber depth was set to 0.234~cm~\cite{Liu1998Twenty}. The optical axial length was set to 1.631~cm~\cite{Liu1998Twenty}. We set the semi-axis $b$ as half the length on the optical axis between the retina and the portion of the lens inside the vitreous chamber. The semi-axis $b$ was obtained by using the lens thickness (with the assumption that 1/3 of the lens thickness in outside of the vitreous), the anterior chamber depth, and the optical axial length:
    \begin{equation*}
        b = \frac{1.631 - (0.234 + 0.66/3)}{2} \text{ cm} = 0.588 \text{ cm}.
    \end{equation*}
    We defined the range for the semi-axis $b$ using the minimum and maximum values for the axial length, the anterior chamber depth, and the lens thickness. Liu et al. \cite{Liu1998Twenty} measured a range of [0.230, 0.253] cm for the anterior chamber depth of rabbits, and a range of [1.618, 1.672] cm for the axial length. From these results, we obtained the lower and greater bounds for the range of the semi-axis $b$:
    \begin{align*}
        b_\text{min} &= \frac{1.618 - (0.253 + 0.697/3)}{2} \text{ cm} = 0.566 \text{ cm}, \\
        b_\text{min} &= \frac{1.672 - (0.230 + 0.66/3)}{2} \text{ cm} = 0.611 \text{ cm}.
    \end{align*}
    \item The vitreous diameter was set to 1.8 cm, using the mean value measured in \cite{Sawada2002Magnetic}, and its range determined from using the standard deviation from the mean reported in this citation, rendering a semi-axis estimate of $a = 0.90$ cm. 
    \item The height of the vitreous-aqueous interface was set to 0.238~cm, so that the ratio of the surface of the vitreous-aqueous humour interface, with the total surface of the ellipse approximately 23\%~\cite{Hutton2016Mechanistic}.
\end{itemize}
The model's geometry for the rabbit had a vitreous volume of 1.7078~ml, which fell within the range of vitreous volumes measured in the literature (1.15-1.8~ml). The constructed geometry had a retinal surface area of 5.4367~cm², which was within the range of 4 to 6 ml measured experimentally~\cite{Reichenbach1991Development}. \\

\subsubsection*{Rat}
When possible, we considered measures for adult rats (120 days-old or older) to inform the model's construction. 
\begin{itemize}
    \item We assumed that the lens was almost entirely immersed in the vitreous chamber cavity, with only a small cap emerging in the anterior chamber. Guided by in situ MRI~\cite{Chui2012Refractive}, we applied a translation of length $l_T/4$ of the lens centre towards the centre of the vitreous ellipse to achieve a similar geometry, so that a small cap of the lens emerged from the vitreous chamber. For simplicity, in order to define the parameter values to construct the geometry, we considered the lens thickness to be entirely inside the vitreous chamber.
    \item The lens thickness was set to 0.387 cm, the mean from \cite{Massof1972Revision}, and its range was set using the mean measurements from \cite{Hughes1979Schematic,Lozano2013Development}.
    \item The lens diameter was set to 0.432 cm, the mean from \cite{Massof1972Revision}, and its range was set using the mean measurements from \cite{Hughes1979Schematic, Pe'er1996Epithelial}.
    \item The optical axial length was set to 0.572~cm, taking the axial length from \cite{Hughes1979Schematic} without the corneal, retina, choroid and scleral thickness measures. The anterior chamber depth was set to 0.062 cm~\cite{Hughes1979Schematic}. As we assumed that the lens was entirely inside the vitreous chamber, we did not need to subtract a portion of the lens thickness from the axial length (as we did for the previous species). This yielded a semi-axis of length
    \begin{equation*}
        b = \frac{0.572 - 0.062}{2} \text{ cm} = 0.255 \text{ cm.}
    \end{equation*}
    We defined the range of values for $b$ using the standard deviation identified for the axial length in \cite{Hughes1979Schematic}.
    \item  The vitreous diameter was set to 0.579 cm, taking the measure of the eye width from \cite{Hughes1979Schematic}, and subtracting from it the retinal, choroid and scleral thickness on both sides of the diameter. This yielded a semi-axis of length $a=0.2895$ cm in the model. We obtained the range of values for $a$ by taking the standard deviation of the vitreous diameter reported in \cite{Hughes1979Schematic}.
    \item The height of the vitreous-aqueous interface was initially determined by fitting our model to the in vivo MRI of a rat~\cite{Chui2012Refractive}, using the visible ciliary body as the end of the retina, which suggested $h_{va} = 0.08$ cm. This yielded a surface ratio of 27.42\% for the vitreous-aqueous interface over the total area of the vitreous ellipsoid, and a retinal surface area of 0.64813~cm². As the retinal surface area we obtained was less than the estimated areas from the literature (ranging from 0.65~cm² to 0.8~cm²~\cite{Mayhew1997Photoreceptor, Baden2020Understanding}), we set $h_{va} = 0.07$~cm, to have a retinal surface area of 0.667~cm². Doing this, we had a retinal surface area that fell inside the range of values identified from the literature, and a model that visually matched the in situ MRI.
\end{itemize}
The model's geometry had a vitreous volume of 51.827 \textmu l, which was close to the vitreous volume of 52.4 \textmu l ($\pm 1.9$ \textmu l) estimated for 120 day-old rats~\cite{Sha2006Postnatal}. The retinal surface area also lay within the literature range, as it was used to define $h_{va}$. \\

\subsubsection*{Mouse}
Given experimental data for murine eyes as a function of age, we chose to consider measures for mice of approximately 3 months old. This was guided by the aim to have a model to compare with experimental results from 8-week-old mice~\cite{Bussing2023Pharmacokinetics}, and constrained by the availability of measurements in the literature. All ocular dimensions considered were measured on mice of strain C57/BL6.
\begin{itemize}
    \item Similar to the rat, the mouse lens is almost entirely situated in the vitreous chamber cavity, with only a small cap emerging in the anterior chamber. Guided by in situ MRI~\cite{Kaplan2010Vitreous, Pan2023Age, Schmucker2004Vivo, Tkatchenko2010Analysis}, we applied a translation of a distance $l_T/4$ of the lens' centre towards the centre of the vitreous ellipse to achieve a similar geometry, so that a small cap of the lens emerged from the vitreous chamber. For simplicity, in order to define the rest of the parameters to construct the geometry, we supposed that the lens thickness was entirely inside the vitreous chamber.
    \item The vitreous chamber diameter was set to 0.3236 cm ($a=0.1618$ cm), taking the mean vitreous chamber diameter for mice aged 89 days~\cite{Tkatchenko2010Analysis}.
    \item Inferred from the linear regression and data points for 3-month-old mice from \cite{Schmucker2004Vivo}, the anterior chamber depth was set to 0.03623 cm and the axial length was set to 0.30727 cm (based on the axial length measure, from which we subtracted the corresponding retinal thickness). Supposing that the entire lens thickness was within the vitreous body, we obtained the semi-axis $b$ by subtracting the anterior chamber depth from the axial length:
    \begin{align*}
        b = \frac{0.30727 - 0.03623}{2} = 0.13552 \text{ cm}.
    \end{align*}
    \item We set the lens diameter and thickness by slightly adjusting the values found in the literature to fit the lens volume to $6.50 \, \mu$l for 3-month-old mice~\cite{Pan2023Age}. As there was a discrepancy between the volume and the measure of the lens' axes in our calculations, we decided to use the volume as reference, as it led to the best visual match with the in situ MRI~\cite{Tkatchenko2010Analysis}. It was reported that mice had lens diameters of approximately 0.225 cm for 3-month-old mice, and lens thicknesses of approximately 0.198 cm~\cite{Pan2023Age}. We incrementally increased these values until we obtained a lens volume close to the one found in the literature, with the constraint that the lens thickness should be less than the lens diameter. We obtained:
    \begin{align*}
        l_D &= 0.240 \text{ cm} \\
        l_T &= 0.216 \text{ cm}.
    \end{align*}
    \item A first attempt to define the height of the vitreous-aqueous interface was made by fitting our model to in vivo MRI (Figure 4B from \cite{Schmucker2004Vivo}), and resulted in $h_{va} = 0.04$ cm. This corresponded to a surface ratio of 25\% for the vitreous-aqueous interface (compared to the total surface of the vitreous chamber ellipsoid), and a retinal surface area of 0.199 cm². As the retinal surface area exceeded the range of measurements found in the literature, we incrementally increased $h_{va}$ until $h_{va} = 0.05$ cm, which yielded a retinal surface area of $A_\text{ret} = 0.188$ cm².
\end{itemize}
The model's geometry had a vitreous volume of 8.42 \textmu l, which was in the range of the vitreous volume measurements from the literature, spanning 4.4 to 12 \textmu l. As mentioned, the retinal surface area measurements from the literature was used to refine the geometry by adjusting $h_{va}$, so surface area comparisons are not feasible.\\
\\

\subsection{Details on the construction of the ensemble of human eye geometries}
We used the data and the results of experimental studies to build an ensemble of human eye dimensions. In most cases, we used the axial length and the vitreous volume measures to reconstruct the eyes, under the assumption of constant anterior chamber depth and lens thickness, and assuming that the eye is axisymmetric around the optical axis. We considered the assumption of a constant anterior chamber depth to be reasonable, based on a weak correlation between the anterior chamber depth and the axial length~\cite{Zhou2020Quantitative}, and based on the high individual variability of the anterior chamber depth between individuals within the same refractive error group~\cite{Fontana1980Volume}. While a correlation has been identified between the lens thickness and the axial length~\cite{Osuobeni1999Ocular}, the reported variability of the lens thickness associated with the axial length is no greater than observed variations of lens thickness found in the population in general (regardless of axial lengths), for example in relation to lens thickness variation with age~\cite{Rosen2006Vitro}. Regardless, by varying the axial length, we obtained a range of eye dimensions covering the variability for the lens thickness and anterior chamber depth.\\
\\
In all cases, we used the same method as described in Section~\ref{sec:SI_details_geom_species} for the human eye to obtain a value of $b$ from the axial length measurement. When no measurement for the vitreous diameter was provided, we used the provided vitreous volume to obtain $a$, with the assumption that the volume of the vitreous chamber ellipsoid formed by $a$ and $b$ is the combination of the vitreous volume and half of the lens volume. The different sources used different measurement and estimation methods, which are summarised in Table~\ref{tab:set_human_eye_description_source_method}.\\
\\
We directly used the measurements from Atchison et al.~\cite{Atchison2004Eye}. From their Table 1, we took the average measurement for the height (vitreous diameter measured in the sagittal plane) and the width (vitreous diameter measured in the axial plane) as the vitreous diameter to obtain $a$, and we took the average length between the axial and sagittal image for the axial length to obtain $b$. We used the digitised measurements of axial lengths and vitreous volumes from the figures presented in Azhdam et al.~\cite{Azhdam2020Vivo}, de~Santana~et~al.~\cite{deSantana2021Use}, and Zhou et al.~\cite{Zhou2020Quantitative} to build the rest of the eye geometries. For Zhou et al.~\cite{Zhou2020Quantitative}, we only kept the data for pathological myopia, as there may be a discrepancy between the figure for emmetropic axial length and volume (Figure 2~\cite{Zhou2020Quantitative}) and their mean and slope specified in the main text (in section 3.3~\cite{Zhou2020Quantitative}). After digitising the data and taking the mean measurements available from Atchison et al. (2004)~\cite{Atchison2004Eye}, we obtained an ensemble of 155 human eye models.\\
\\
\begin{landscape}
\begin{table}[]
    \centering
    \begin{tabular}{|p{2.5cm} | p{3cm} | p{5cm}| p{2.5cm} | p{3cm} | p{5cm} |}
    \hline
        Source & Sample size & Inclusion criteria & Refractive error range & Measurement of AL & Measurement of volume (or other measurements)  \\ \hline
        Atchison et al. (2004) \cite{Atchison2004Eye} & 88 eyes & Aged between 18 and 36 years, good ocular health. & Emmetropic and myopic. & MRI & Other ocular measurements (height, width of vitreous volumes) were estimated by MRI. \\ \hline
        Azhdam et al. (2020) \cite{Azhdam2020Vivo} & 100 eyes & No ocular pathology and no history of ocular surgery. & Not specified. & Estimated from CT scan. & Volume estimated by CT scan with Mimics image analysis tool to estimate vitreous volume. \\ \hline
        Zhou et al. (2020) \cite{Zhou2020Quantitative} & 290 eyes & Met the diagnostic criteria of pathological myopia, aged between 18 and 60 years, no history of ocular diseases affecting diopter, and no history of ocular surgery. & Pathological myopia. & Optical biometry. & Volume estimated by MRI.\\ \hline
        de Santana et al. (2021) \cite{deSantana2021Use} &  112 eyes & Pseudophakia, aged older 50 years, eyes with axial length between 2.1 to 2.6 cm. & Not specified. & Optical biometry & Vitrectomy and, after the fluid-air exchange, the vitreous chamber was filled with a dye. The infused volume of each eye was recorded. \\ \hline
    \end{tabular}
    \caption{Description of sources, methods, and measurements used to build the ensemble of human eyes.}
    \label{tab:set_human_eye_description_source_method}
\end{table}
\end{landscape}

\section{Results of the global sensitivity analysis on the MFPT}
\begin{figure}[h]
        \centering
        \includegraphics[width=0.7\linewidth]{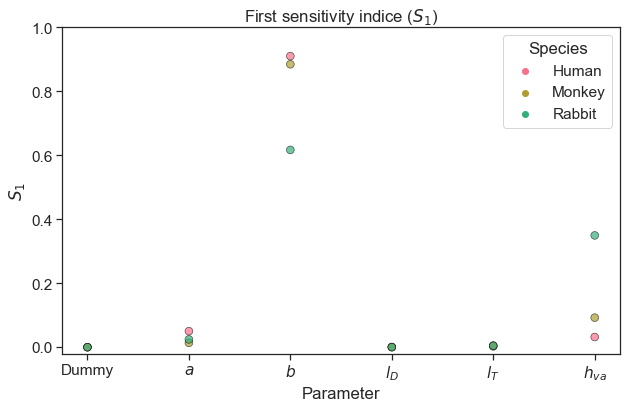}
    \caption{Solution of the global sensitivity analysis of the MFPT for a Fab molecule for an injection location at $P_m$, for the human, cynomolgus monkey, and rabbit eye models, and with varying geometrical parameters (within their identified uncertainty range in Table 3). The semi-axes $a$ and $b$ are the semi-major and semi-minor axis of the vitreous chamber ellipse, $l_D$ and $l_T$ are the lens diameter and thickness, and $h_{va}$ is the height of the vitreous-aqueous humour interface, as defined in Figure 1 of the main text.}
    \label{fig:global_SA}
\end{figure}

Figure~\ref{fig:global_SA} shows the results of the global sensitivity analysis performed to analyse the impact of the geometrical parameters on the MFPT for a Fab molecule format, injected at the injection point $P_m$. In the human, cynomolgus monkey and rabbit eye, $P_m$ corresponds to the midpoint of the vitreous chamber depth along the optical axis, and in the rat and mouse eye, $P_m$ corresponds to the midpoint between the retina and the lens along the vitreous diameter (see Figure 2). The total sensitivity indices are not illustrated, as they do not differ to the first sensitivity indices. The global sensitivity analysis identified that, within the uncertainty range of each parameter and for an injection at $P_m$, the length of the semi-axis $b$, as depicted in Figure 1 of the main text, was the most sensitive for the MFPT.\\
\\

\section{Additional figures}
\subsection{Results of the ensemble of human eye models, excluding pathologically myopic eyes}
\begin{figure}
    \centering
    \includegraphics[width=\linewidth]{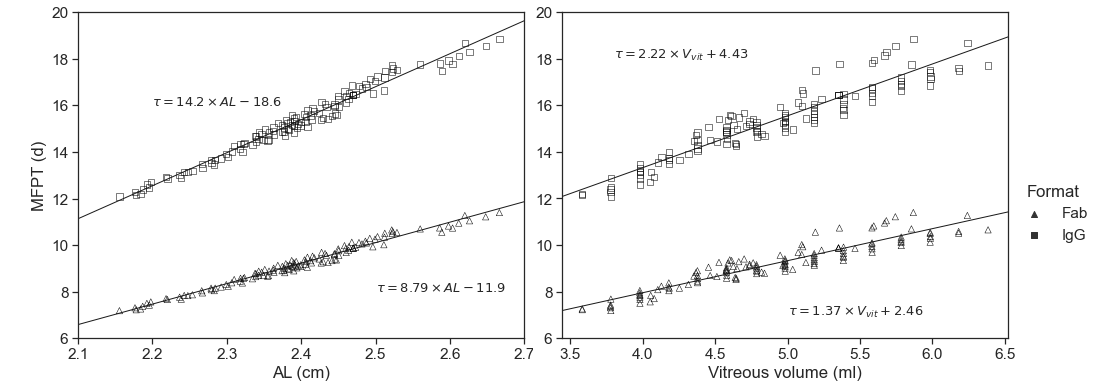}
    \caption{Numerical solution and linear regressions of the MFPT for an injection at $P_m$, for different molecular formats and with parameters defined in Table 1, for the ensemble of human eye models without pathology, plotted against the axial length (AL) and the vitreous volume. }
    \label{fig:mfpt_ensemble_human_eye_without_pathology}
\end{figure}

Figure~\ref{fig:mfpt_ensemble_human_eye_without_pathology} shows the MFPT in the ensemble of human eyes without the pathological myopia dataset, plotted against the axial length (AL) and the vitreous volume.\\
\\

\subsection{Conditional MFPT}
\begin{figure}
    \centering
    \begin{subfigure}{\linewidth}
        \centering
        \includegraphics[width=0.9\linewidth]{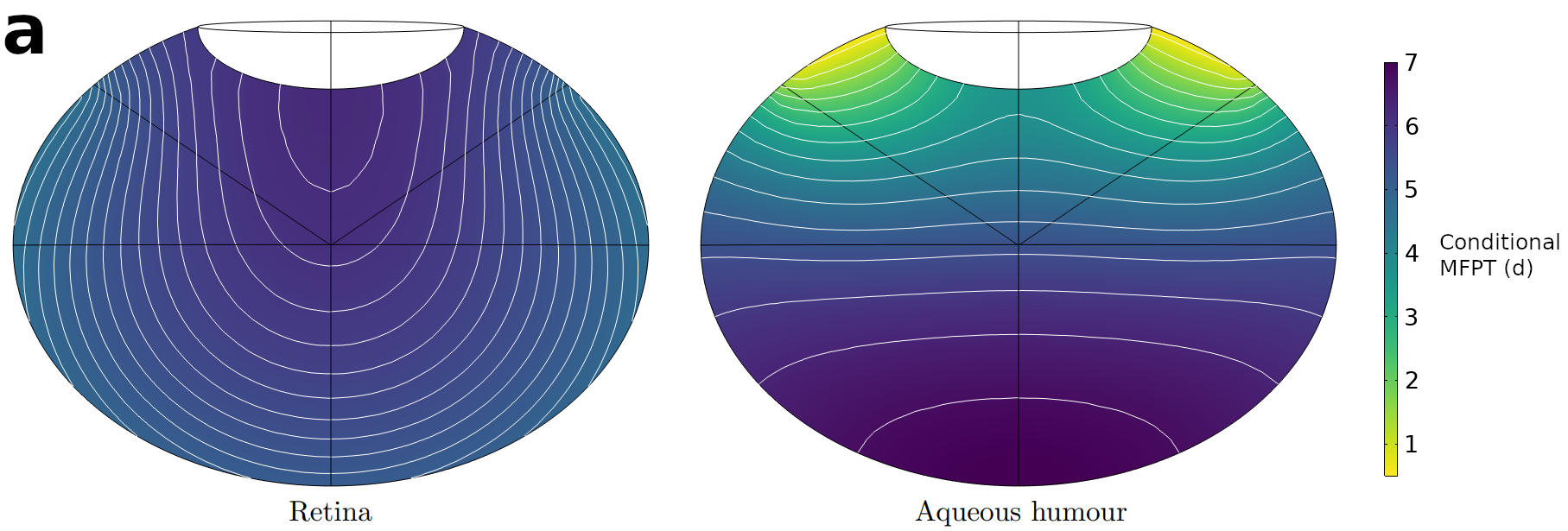}
    \end{subfigure}
    \\
    \begin{subfigure}{\linewidth}
        \centering
        \includegraphics[width=0.9\linewidth]{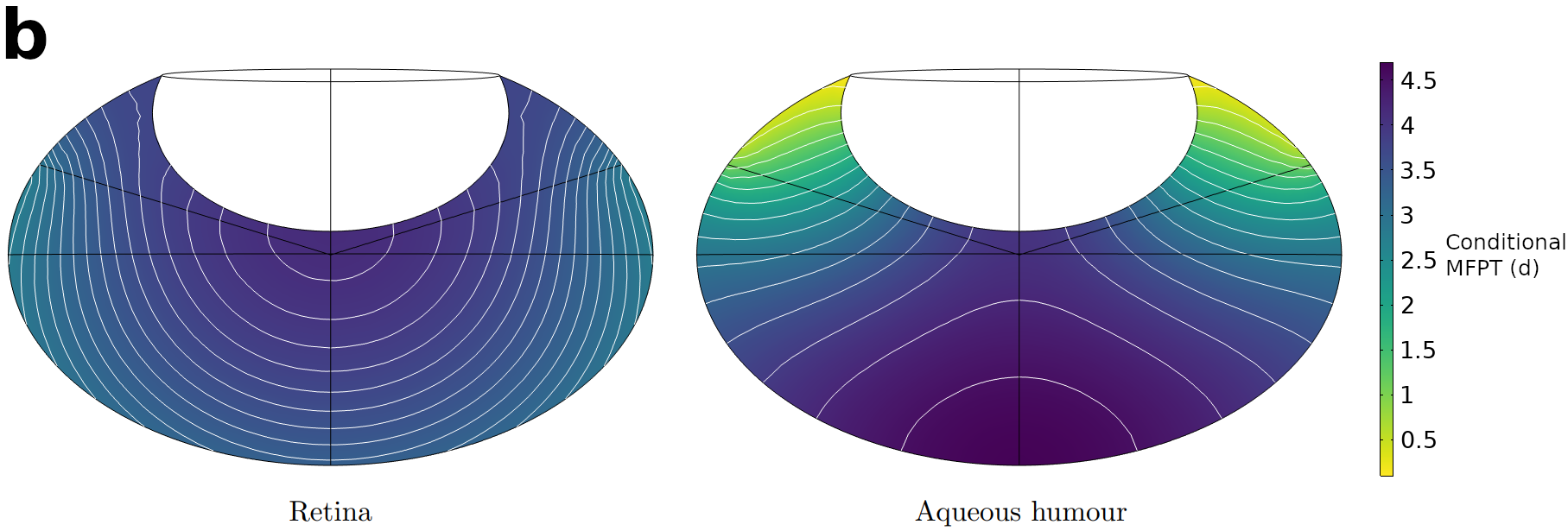}
    \end{subfigure}
    \\
    \begin{subfigure}{\linewidth}
        \centering
        \includegraphics[width=0.9\linewidth]{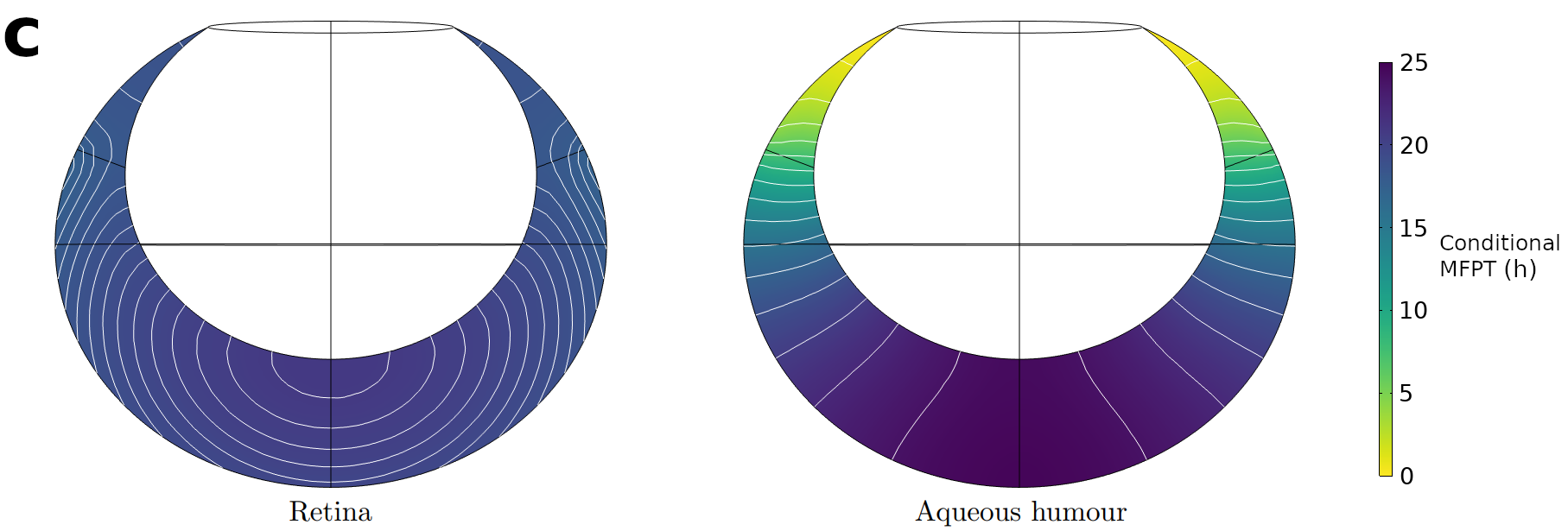}
    \end{subfigure}
    \\
    \begin{subfigure}{\linewidth}
        \centering
        \includegraphics[width=0.9\linewidth]{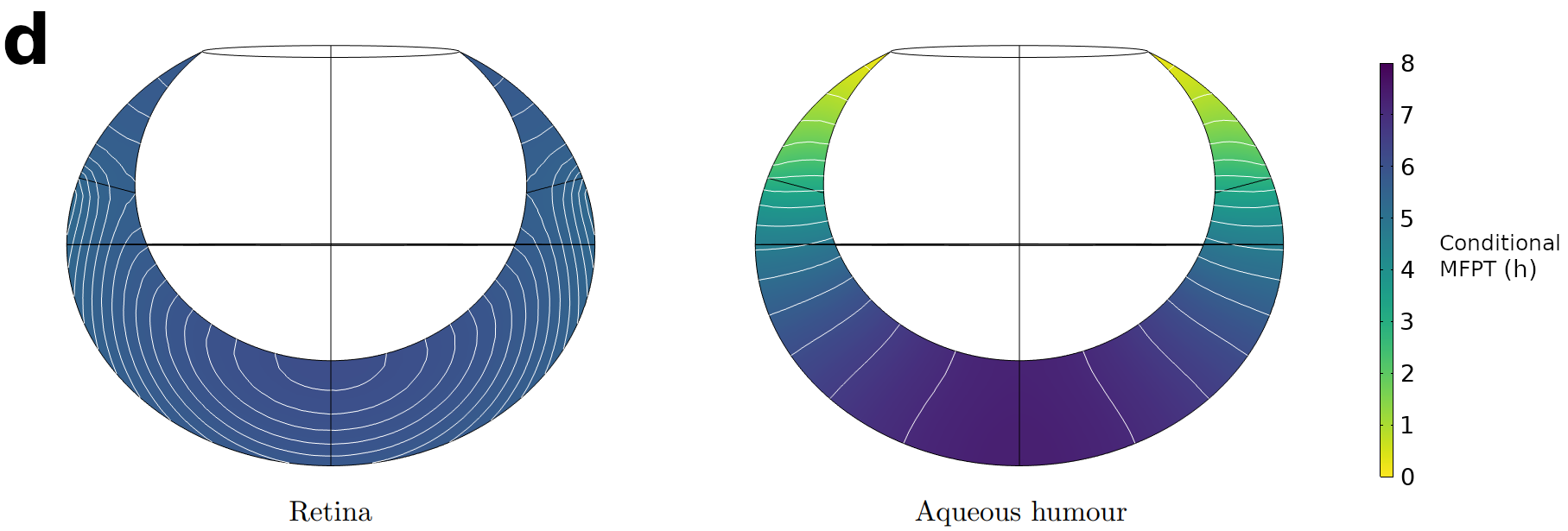}
    \end{subfigure}
    \caption{Numerical solution and contour lines for the MFPT, conditional on exiting through the vitreous-retina and vitreous-aqueous humour interfaces for a Fab molecule as a function of injection site, for a) cynomolgus monkey, b) rabbit, c) rat and d) mouse eye models. The parameters for these plots are in Table 1 and Table 3, and the geometries used are illustrated in Figure 2.}
    \label{fig:SI_cond_MFPT}
\end{figure}

To obtain numerical solutions for the conditional MFPT, equations (2.3) and (2.4) were solved with parameter values given in Table 1 for a Fab molecule, using the eye geometry for the cynomolgus monkey, rabbit, rat and mouse (Figure 2). Figure~\ref{fig:SI_cond_MFPT} shows the results for the conditional MFPT for particles exiting through the vitreous-retina and vitreous-aqueous humour interfaces.

\newpage
\begin{singlespace}
\bibliography{paper_biblio}
\end{singlespace}